# Recommending on graphs: a comprehensive review from a data perspective

**Lemei Zhang**[1] · **Peng Liu**[1] · **Jon Atle Gulla**[1]

**Abstract**
Recent advances in graph-based learning approaches have demonstrated their effectiveness in modelling users' preferences and items' characteristics for Recommender Systems (RSs). Most of the data in RSs can be organized into graphs where various objects (e.g. users, items, and attributes) are explicitly or implicitly connected and influence each other via various relations. Such a graph-based organization brings benefits to exploiting potential properties in graph learning (e.g. random walk and network embedding) techniques to enrich the representations of the user and item nodes, which is an essential factor for successful recommendations. In this paper, we provide a comprehensive survey of Graph Learning-based Recommender Systems (GLRSs). Specifically, we start from a data-driven perspective to systematically categorize various graphs in GLRSs and analyse their characteristics. Then, we discuss the state-of-the-art frameworks with a focus on the graph learning module and how they address practical recommendation challenges such as scalability, fairness, diversity, explainability, and so on. Finally, we share some potential research directions in this rapidly growing area.

## 1 Introduction

In the last few decades, the rapid development of Web 2.0 and smart mobile devices has resulted in the dramatic proliferation of online unstructured data, such as news articles.

✉ Lemei Zhang
lemei.zhang@ntnu.no

Peng Liu
peng.liu@ntnu.no

Jon Atle Gulla
jon.atle.gulla@ntnu.no

[1] Department of Computer Science, NTNU, 7491 Trondheim, Norway

They are explicitly or implicitly connected with each other and can naturally be formed into graphs representing objects and their relationships in varied domains, including e-commerce, social networks, and so on. On the one hand, the interconnection of objects shows a direct (e.g. social relations in a social network) or indirect interactive relationship (e.g. item co-occurrence in an item homogeneous network), which provides a more intuitive and effective way for recommendation systems to explore the hidden relationships between the target user and the recommended items. On the other hand, the data structure of graphs breaks the independent interaction assumption[1] by linking users or items with their associated attributes such that the recommender systems are able to capture not only the user–item interactions but also the rich underlying connections by mining item–item/user–user relations to make more accurate recommendations. Moreover, most recommendation models work as black boxes that only provide predictive results rather than exhibiting the reasons behind a recommendation, such as collaborative signals in collaborative filtering or knowledge-aware reasoning in knowledge graph-based recommendation. Some recommender systems give users such as "users who bought A also bought B" as an explanation of the recommended results. However, no in-depth explanations of the intrinsic recommendation mechanism for the selected items may result in the users over-relying on the recommendation system and ignoring that the purpose of the recommendation is to make the recommendation platform profitable (Jannach et al. 2019). Recent advances in graph-based recommender systems have demonstrated their effectiveness in improving the explainability of the recommender systems by using explicit connections between objects in graphs to reveal the recommendation results (He et al. 2015; Ma et al. 2019; Hu et al. 2018). Therefore, it is of crucial significance to fully explore the semantic connections and potential relations of the graphs to improve the performance on both the explainability and accuracy of recommendations.

However, there exist some problems and challenges in how recommender systems (RSs) can make full use of these data:

(1) Heterogeneous Objects: Unstructured data can be organized into a graph including different-typed objects and links. Modelling and abstracting such a space of information have been a challenging task encountered in RSs.
(2) Large-scale Volume: Real graphs, such as social networks, can easily have millions even billions of nodes and edges, which renders most traditional recommendation algorithms computationally infeasible.
(3) Dynamic Contents: Most real-world graphs are intrinsically dynamic with addition/deletion of edges and nodes. Meanwhile, similar to a graph structure, node attributes also change naturally such that new content patterns may emerge and outdated content patterns will fade.

Recently, graph learning (GL) has exhibited the potential to obtain knowledge embedded in different kinds of graphs. Many GL techniques, such as random walk, graph embedding, and graph neural networks, have been developed to learn the complex relations modelled on graphs and achieve a great improvement in recommendation performance. An emerging RS paradigm built on GL, namely Graph Learning-based Recommender Systems (GLRSs), has attracted growing attention in both research and industry communities. For example, researchers leverage random walk to propagate users' preference scores from historical item nodes and output a preference distribution over unobserved items, such as ItemRank (Gori et al. 2007) over the item–item correlation graph, RecWalk (Nikolakopoulos and Karypis 2019) over the user–item bipartite graph, and TriRank (He et al. 2015) over the user–item–aspect tripartite graph. Moreover, various graph embedding techniques and graph neural networks have been proposed and incorporated into the representation learning of RSs, using direct or multi-hop connections within graphs to enrich the representations of the user and item nodes. These approaches further improve the recommendation performance.

[1] The techniques, e.g. factorization machine (Koren et al. 2009), Neural FM (He and Chua 2017), based upon the independent interaction assumption model the user or item as a single object, but ignore the connection relationship among them.

To date, there are only a handful of literature reviews related to our paper. Wang et al. (2021a) survey KG-embedding models for link prediction. In Guo et al. (2020), Liu and Duan (2021), the authors summarize recent works utilizing KGs as side information for accurate and explainable recommendations. The authors of (Wu et al. 2020a) summarize the most recent works on GNN-based recommender systems and propose a classification schema for organizing existing works. Our work differs from the previous works in that we give a systematic and comprehensive review of recommendation techniques starting with different types of data-driven graphs rather than focusing on one specific branch.

Another closely related topic is linked data-based recommendation. Figueroa et al. (2015) present a systematic literature review to summarize the state of the art in RS that uses structured data published as Linked Data for providing recommendations of items from diverse domains. Tarus et al. (2018) present a comprehensive review of ontology-based recommendations for e-learning. The main distinction is that these works discuss making recommendations with the subject of linked data, which is structured interlinked data that manifest as a Web of Data from multiple sources. [2] In fact, both works focus on knowledge-based recommendations without referring to recent deep learning-based technologies. In this paper, we focus on how raw data can be extracted into a wider range of graphs (e.g. tree-based graphs, homogeneous graphs, and hypergraphs) and how traditional as well as state-of-the-art graph learning techniques can be applied to these graphs for recommendation purposes.

Although there is a variety of literature on this subject, only one study analysed the role of GLRSs (Wang et al. 2020e). The limitations lie in that they only cover a limited number of references and do not go into the technical details on graph learning modules in GLRS and how they address those challenges. Furthermore, they do not give a systematic summarization of existing datasets adopted by graph learning-based recommendation research. To overcome such an information gap, in this paper, we contribute the most comprehensive overview of state-of-the-art GLRSs. We systematically analyse the benchmark datasets in GLRSs, provide detailed descriptions of representative models, make the necessary comparisons, and discuss their solutions to practical recommendation issues such as scalability, fairness, diversity, and explainability.

[2] https://en.wikipedia.org/wiki/Linked_data.

**Contributions:** This survey provides a thorough literature review on the approaches of graph learning-based recommender systems and the involved various types of graphs from a data perspective. It provides a panorama starting from various data characteristics to applied technologies, with the hope that both academic researchers and industrial practitioners can have a rough guideline and step into the field of graph learning-based recommendation from the data resources available at hand. This survey serves to promote the innovation and development in the field of GLRSs, while exploring the possibility of enhancing the richness by discussing and summarizing existing open issues in the field. To this end, the main contributions of this work are threefold:

- We explore different data input categories based on their acquisition and intrinsic characteristics and further proposed a novel taxonomy to categorize various graphs in GLRSs from data perspectives. Meanwhile, we summarize resources regarding GLRSs, including benchmark datasets and open-source knowledge graphs.
- We conduct a systematic literature review on traditional and recent developments and progress of graph learning-based techniques for recommender systems, which correspond to associated different graph taxonomies.
- We analyse the limitations of existing works and suggest future research directions of GLRSs such as dynamicity, interpretability, and fairness, for giving references for this community.

**Organization of Our Survey:** The rest of the survey is organized as follows. In Sect. 2, we review our research methodology on how we collected the related papers and provide an initial analysis of datasets adopted by reference papers. In Sect. 3, we introduce the definitions of the basic concepts required to understand the graph learning-based recommendation problem, followed by a formal problem definition of graph learning-based recommendation. Section 4 provides a new taxonomy of graphs that are related to specific datasets. Section 5 provides an overview of state-of-the-art GLRSs techniques. Section 6 discusses the current challenges and suggests future directions, followed by the conclusions in Sect. 7.

## 2 Research methodology

### 2.1 Paper collection

To achieve a systematic structure of existing research on graph-based recommendation, this study was performed based on a bibliographic review proposed by Webster and Watson (2002), Kitchenham (2004), Wolfswinkel et al. (2013). Specifically, we first conducted a comprehensive review of previously published papers concerning GLRSs and used Scopus as the main source of information. The search strings are listed in Table 1. Other bibliographic databases and archives also constitute the auxiliary sources used for literature search, such as ACM Digital Library,[3] IEEE Xplore, [4]

[3] https://dl.acm.org.

[4] https://ieeexplore.ieee.org.

**Table 1** Search string

| Key words | Search string |
|---|---|
| graph, hypergraph, recommender, recommendation, recommender system, recommendation system, recommendation service, recommendation approach, recommendation model, recommendation method, recommendation algorithm, recommendation application, recommendation engine, recommendation agent, recommendation framework, collaborative filtering, social recommendation, representation learning, knowledge graph, graph neural network | (TITLE-ABS-KEY*(recommender) OR TITLE-ABS-KEY(recommendation system) OR TITLE-ABS-KEY(recommendation service) OR TITLE-ABS-KEY(recommendation approach) OR TITLE-ABS-KEY(recommendation model) OR TITLE-ABS-KEY(recommendation method) OR TITLE-ABS-KEY(recommendation algorithm) OR TITLE-ABS-KEY(recommendation application) OR TITLE-ABS-KEY(recommendation engine) OR TITLE-ABS-KEY(recommendation agent) OR TITLE-ABS-KEY(recommendation framework) OR TITLE-ABS-KEY(collaborative filtering) OR TITLE-ABS-KEY(social recommendations)) **AND** (TITLE-ABS-KEY(graph) OR TITLE-ABS-KEY(hypergraph) OR TITLE-ABS-KEY(knowledge graph) OR TITLE-ABS-KEY(graph neural network) OR TITLE-ABS-KEY(representation learning)) |

*TITLE-ABS-KEY is a combined field that searches abstracts, keywords, and document titles

Springer, [5] ResearchGate, [6] and Web of Science. [7] We conducted the same keyword-based search in these search engines.

We first checked the paper titles and then reviewed the abstracts, keywords, results, and conclusions to obtain the first list of studies. We then double-checked the reference list in those papers to identify additional studies that were relevant to our review topic. After that, the publications retrieved needed to be further filtered in order to eliminate false positives, which are irrelevant to the current survey. Therefore, a pre-defined set of inclusion and exclusion criteria displayed in Table 2 were applied to the retrieved papers. Finally, we obtained a collection of 182 papers that meet the mentioned criteria and then are summarized in Sects. 3 and 4. Figure 1 gives the statistics of the collected papers with the publication time and venue.

Note that these papers were mainly selected according to the two criteria: (1) publication time; (2) impact. Therefore, the references cited in this paper on the field of GLRS are representative, but still limited. The uncited literature is only limited by the length of this paper and the pre-set filtering criteria.

5 https://www.springer.com.

6 https://www.researchgate.net/.

7 https://www.webofknowledge.com.

**Table 2** Inclusion criteria and exclusion Criteria

| Criteria | Inclusion Criteria | Exclusion Criteria |
|---|---|---|
| Recommender System | The study focused on recommender system in multiple domains | The study presents a system or technique other than a recommender system |
| The use of graph as input | The study presents a system that uses a type of graph as the input, and other data structures such as texts, image or acoustic information can also be the auxiliary means of system input | The study presents a system using a data structure except graph as input |
| Publication Date | The paper is published between 2007 and September 2021 | The paper is published before 2007 or after September 2021 |
| Language | The paper is written in English | The paper is written in a language different than English |
| Publication type | The paper has been peer reviewed and published in prestigious and top-tier international conferences and journals e.g. SIGIR, NIPS, ICML, RecSys, CIKM, ICLR, AAAI, IJCAI, WWW, WSDM, KDD, UMAP, TOIS, TKDE, and UMUAI etc. The paper is a primary study | The paper has not been peer reviewed (e.g. theses, books, technical reports, (extended) abstracts, talks, presentations, tutorials, guidelines) or not published in top-tier conferences or journals. The paper is a secondary study (e.g. systematic literature review, survey) |
| Accessibility | The paper's content can be accessed from a technical university (e.g. Norwegian University of Science and Technology) without additional payment | The paper's content cannot be accessed from a technical university (e.g. Norwegian University of Science and Technology) without additional payment |

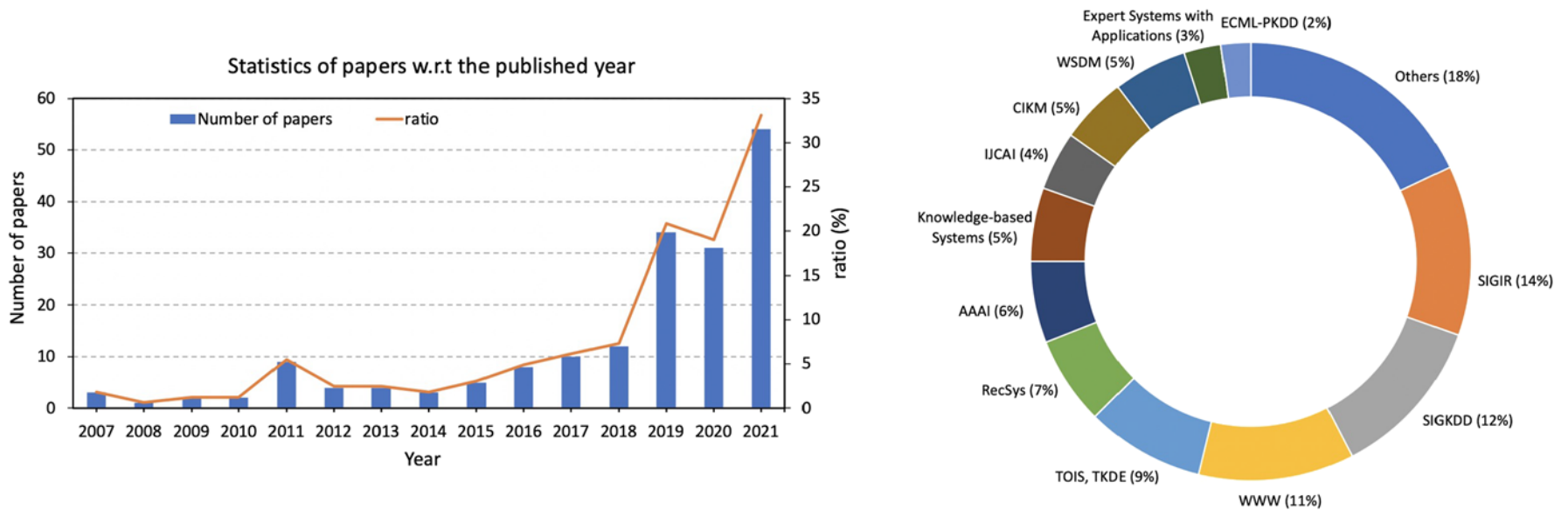

**Fig. 1** Statistics of publications related to GLRSs grouped by the publication year and venue

### 2.2 Data analysis

Analyzing the collected papers, we made two observations on the utilized datasets: (1) They were across different domains, such as e-commerce and entertainment domains; (2) Some datasets could be used to construct multiple types of graphs for different recommendation purposes, while some were only used to construct one type of graph. For instance, we found nearly all classified graph types were utilized for the Amazon dataset, while only a multi-source graph could be found for the Epinions dataset. To clearly make comparisons and show the difference of these datasets in terms of both domains and graph types, we made a detailed comparison of all datasets in Table 5 ("Appendix A").

Recent advances in GLRS either focus on incorporating explicit or implicit user/item information into the process of mapping, and design learning algorithms for specific graph structures (Christoforidis et al. 2021; Pang et al. 2022), or focus on how to incorporate time information into graph forming and learning processes to better capture the dynamic needs of users to improve recommendation performance (Fan et al. 2021; Zhang et al. 2022). In Lv et al. (2021), the authors make reproductions of 12 heterogeneous graph neural network based modes and test them on 11 public available datasets regarding node classification, link prediction and knowledge-aware recommendation tasks. Their reported experimental results reveal that the superiority of the performance of the most advanced HGNN models rely on the lack of fair comparison with the homogeneous GNNs and other baselines in their original paper. Beside, some works also reveal issues such as data leakage, tuning on test set, cost of large amount of memory, and time for training. Meanwhile, the authors of Lv et al. (2021) release a heterogeneous graph benchmark (HGB) for open, reproducible heterogeneous graph research, and present a GAT-based heterogeneous GNN model resulting in promising results on three aforementioned tasks.

We count the corresponding datasets for GLRS technologies adopted in this survey and present the results in Table 7 ("Appendix C"). From the table, we can observe that both datasets and their leveraged GLRS technologies are distributed extremely unbalanced. The reason for the former one may be attributed to whether the dataset is public, the date of publication, whether the dataset contains various attributes of users/items and so forth. The reason for the uneven distribution of technology can be roughly attributed to the development of deep learning in the field of graph learning to bring more possibilities.

## 3 Problem formalization

In this section, we first introduce the definition of the basic concepts in graph-based recommendations and then provide a formal definition of the graph-based recommendation problem.

### 3.1 Basic definitions

The definitions related to GLRSs are as follows.

**Definition 1** (*Graph*) A graph is $\mathcal{G} = (V, E)$, where $v \in V$ is a node and $e \in E$ is an edge. Each edge $e_{ij}$ is a pair between vertex $v_i$ and $v_j$. Each edge of $\mathcal{G}$ can be mapped to a real number (if any), denoted as $W : E \rightarrow R^+$ from edge $e \in E$ to a real number $w \in R^+$. Such weights can represent e.g. costs, lengths or capabilities, depending on the specific problem. $\mathcal{G}$ is associated with a node type mapping function $f_v : V \rightarrow \mathcal{A}$ and an edge type mapping function $f_e : E \rightarrow \mathcal{R}$. $\mathcal{A}$ and $\mathcal{R}$ denote the set of node types and edge types, respectively. Each node $v_i \in V$ belongs to one particular type, i.e. $f_v(v_i) \in \mathcal{A}$. Similarly, for $e_{ij} \in E$, $f_e(e_{ij}) \in \mathcal{R}$. When a graph has $e_{ij} \not\equiv e_{ji}$ and $f_e(e_{ij}) \not\equiv f_e(e_{ji})$, it is a *directed graph*. Otherwise, the graph is *undirected*.

**Definition 2** (*Network Schema*) The network schema (Sun and Han 2013), denoted as $T_{\mathcal{G}} = (\mathcal{A}, \mathcal{R})$, is a meta template for a heterogeneous network $\mathcal{G} = (V, E)$ with the node type mapping $f_v$ and the edge mapping $f_e$, which is a directed graph defined over node type set $\mathcal{A}$ with edges as relations from $\mathcal{R}$.

**Definition 3** (*Homogeneous Graph*) A homogeneous graph $\mathcal{G}_{homo} = (V, E)$ is a graph in which $|\mathcal{A}| = |\mathcal{R}| = 1$. This is to say that all nodes in $\mathcal{G}$ belong to a single type and all edges to one single type.

**Definition 4** (*Tree Graph*) A tree graph $\mathcal{G}_{tree} = (V, E)$ is a graph in which all nodes are connected with each other and there is no cycles in $\mathcal{G}$. The *leaf* node in a tree graph has degree 1, where *degree* of a vertex $v$, denoted as $d(v)$, is defined as the number of vertices that are adjacent to $v$.

**Definition 5** (*Heterogeneous Graph*) A heterogeneous graph $\mathcal{G}_{heter} = (V, E)$ can be defined as a graph in which $|\mathcal{A}| > 1$ and/or $|\mathcal{R}| > 1$.

**Definition 6** (*k-partite Graph*) A k-partite graph $\mathcal{G}_{kpar} = (V, E)$ is a graph in which nodes are partitioned into $k$ different disjoint sets $\{\mathcal{A}_1, \mathcal{A}_2 \ldots \mathcal{A}_k\}$ where $\mathcal{A} = \mathcal{A}_1 \cup \mathcal{A}_2 \cup \ldots \cup \mathcal{A}_k$ and $\mathcal{A}_i \cap \mathcal{A}_j = \emptyset$, where $i \neq j; i, j \in \{1, \ldots, k\}$. In $\mathcal{G}_{kpar}$, no two nodes within the same set are adjacent. If $\exists e_{mn}$ between nodes $v_m$ and $v_n$ where $f_v(v_m) \in \mathcal{A}_i$ and $f_v(v_n) \in \mathcal{A}_j$, then $i \neq j; i, j \in \{1, \ldots, k\}$.

**Definition 7** (*Knowledge Graph*) A knowledge graph $\mathcal{G}_{know} = (V, E)$ is a directed graph whose nodes are entities and edges are *subject–property–object* triple facts. Each edge of the form (head entity, relation, tail entity), denoted as $< h, r, t >$, indicates a relationship of $r$ from entity $h$ to entity $t$. $h, t \in V$ are entities and $r \in E$ is the relation. Entities and relations in a knowledge graph are usually of different types, such that $\mathcal{A} \cap \mathcal{R} = \emptyset$. Knowledge graphs can be viewed as another type of heterogeneous graphs.

The network schema of a heterogeneous graph specifies type constraints on graph objects/nodes and relationships of links/edges between the objects/nodes. The constraints make the heterogeneous graph semi-structured data, guiding the exploration of the semantics of the graph.

### 3.2 Problem definition

Given a data source $\mathcal{X}$, normally a user set $\mathcal{U}$ and an item set $\mathcal{I}$, for each user $u \in \mathcal{U}$, the recommendation problem can generally be seen as a mapping function $Y = argmax_u f(\mathcal{U}, \mathcal{I})$ generating the corresponding recommendation results from $\mathcal{I}$ that are of interests to the user $u$. However, there is no formal definition of GLRSs to date due to different implementations of various models on different datasets with specific characteristics. Graphs in GLRSs can be built upon input data sources, e.g. user–item interactions as well as other auxiliary information, for instance, considering the graph $\mathcal{G} = (V, E)$, where nodes in $V$ can represent e.g. users, items and other named entities, while edges in $E$ can represent e.g. purchases, clicks, social relations as well as other relationships among entities. In this survey, we formulate the GLRS problem from a general perspective. Specifically, for the input data source(s) $\mathcal{X}$, we would like to find a mapping $\mathcal{M}(\mathcal{X}) \rightarrow \mathcal{G}$, which is used as the input to generate the corresponding recommendation results $Y$ by modelling graph properties as the main way complemented with other auxiliary features of graphs:

$$Y = argmax_u f(\mathcal{M}(\mathcal{X}) \rightarrow \mathcal{G}|\Theta) \tag{1}$$

where $\mathcal{G}$ can be of different types, e.g. homogeneous, k-partite, tree-based, complex heterogeneous graph, etc., based on specific recommendation scenarios, while $Y$ can be of different forms, e.g. rating scores, possible links, classifications, or ranked lists. $\Theta$ is the model parameter set to be optimized during model training. In this survey, we will focus on different types of input data sources $\mathcal{X}$, different types of graphs $\mathcal{G}$ formed from $\mathcal{X}$, the main technologies used for recommendation purposes $f$, and the connections between these aspects to elaborate on GLRSs-related studies.

## 4 From data to graphs

Most of the existing reviews on GLRSs only focus on the input graph types and related recommendation technologies, but none of them associated different graph types with original datasets. In fact, the construction of the graph is largely determined by the dataset at hand. A large amount of semantic and structural information is hidden in the graphs constructed by analyzing the dataset, and more and more research results show that the performance of recommendation can be improved in many aspects, such as accuracy (Wang et al. 2020d; Sun et al. 2021), fairness (Farnadi et al. 2018), diversity (Mansoury et al. 2020; Isufi et al. 2021), and explainability (Liu et al. 2021a), by appropriate learning and modelling the graphs. Based on this, before diving into the specific types of graphs, it is necessary to understand the input data structures, their attributes, and how these attributes are related to the formation of graphs. Broadly speaking, there are two types of input data: user–item interaction and side information (Shi et al. 2014), from which side information can also be further classified into user/item attributes and external resources. Hence, we classify the input data into three major categories: interactive data, context data, and external data. The rationale of the classification is twofold: (1) If it is interactive data for GLRS, e.g. user–item

**Table 3** A summary of the data, corresponding graph types, recommendation tasks and representative techniques in GLRS

| Data Class | Graph Type | Recommendation Task | Representative Technique Category |
|---|---|---|---|
| Interactive Data | Homogeneous graph | Sequential Recommendation, Top-N Recommendation | PRM (Gori et al. 2007), TEM (Wang et al. 2020b), DNN (Xu et al. 2019a; Isufi et al. 2021), GNN (Qiu et al. 2020b), DHM (Wu et al. 2019c; Abugabah et al. 2020) |
| | K-partite graph | Link Prediction, Top-N Recommendation | KML (Li and Chen 2013), RWM (Eksombatchai et al. 2018), DNN (Zheng et al. 2018), AM (Wang et al. 2019d), GNN (Wang et al. 2019f; Chen et al. 2020a) |
| | Hypergraph | Top-N Recommendation | DHM (Wang et al. 2020c) |
| | Multiple graph | Link Prediction | GDRM (Verma et al. 2019) |
| Contextual Data | Tree-based graph | Sequential Recommendation, Rating Prediction | LFM (Koenigstein et al. 2011), GDRM (He et al. 2016), DNN (Huang et al. 2019), AM (Gao et al. 2019b) |
| | Homogeneous graph | Sequential Recommendation, Top-N Recommendation | RWM (Jamali and Ester 2009), TEM (Gao et al. 2018), DHM (Song et al. 2019b; Chen et al. 2019a) |
| | K-partite graph | Next-item Recommendation, Link Prediction, Rating Prediction | MPM (Yu et al. 2014), GDRM (Cen et al. 2019), TEM (He et al. 2015), AE (Zhang et al. 2019a), AM (Xin et al. 2019), GNN (Kim et al. 2019; Ying et al. 2018), Others (KNN) (Phuong et al. 2019) |

**Table 3** continued

| Data Class | Graph Type | Recommendation Task | Representative Technique Category |
|---|---|---|---|
| | Complex heterogeneous graph | Top-N Recommendation, Rating Prediction | LFM (Zhao et al. 2017), RWM (Bagci and Karagoz 2016; Ma et al. 2019), RM (Shi et al. 2020), MPM (Shi et al. 2015), GDRM (Shi et al. 2018; Fu et al. 2020), TEM (Chen et al. 2019d), AE (Zhang et al. 2016), AM (Han et al. 2018), GNN (Wang et al. 2019e), Others (Spectral Clustering) (Farseev et al. 2017) |
| | Hypergraph | Top-N Recommendation | RTGR (Bu et al. 2010), GNN (Yu et al. 2021), DHM (Gharahighehi et al. 2020) |
| | Multiple graph | Top-N Recommendation, Rating Prediction | LFM (Ma et al. 2011a), GDRM (Ali et al. 2020), AM (Vijaikumar et al. 2019), GNN (Fan et al. 2019b; Wu et al. 2019b), DHM (Monti et al. 2017) |
| External Data | K-partite graph | Rating Prediction | RWM (Jiang et al. 2018a), MPM (Lu et al. 2020; Yu et al. 2013) |
| | Complex heterogeneous graph | Top-N Recommendation, Sequential Recommendation, Rating Prediction | PRM (Catherine et al. 2017; Catherine and Cohen 2016), MPM (Ostuni et al. 2013), GDRM (Palumbo et al. 2017; Jiang et al. 2018b), TEM (Wang et al. 2018b, 2019a; Ai et al. 2018), DNN (Sun et al. 2018; Wang et al. 2019g), GNN (Wang et al. 2019c), DRL (Song et al. 2019a; Lei et al. 2020a), DHM (Sheu and Li 2020; Wang et al. 2020g; Zhou et al. 2020a; Wang et al. 2018c) |
| | Multiple graph | Top-N Recommendation | DNN (Wang et al. 2017) |

interactions, or context data on user/item side. (2) Whether the information should be from additional data sources, e.g. external knowledge bases. Table 3 illustrates the taxonomy of datasets, and their relations to graph types as well as techniques in GLRSs.

**Interactive Data in GLRSs.** Recommendation is inheritly a tool and technique that provides users with potentially interesting items based on past user–item interactions. The user–item interaction data as a prerequisite for GLRS naturally form a relational connection between the user and item. The connections can appear in explicit or implicit form depending on whether obvious numerical numbers or positive/negative responses are directly observed. User implicit feedback data are inferred by indirect user behaviour such as clicking, page viewing, purchasing, watching and listening (example datasets are Last.fm (Cantador et al. 2011), Bing-News (Wang et al. 2018b, c), and YahooMusic (Dror et al. 2012)), whereas explicit interactive data are collected directly by prompting users to provide numerical feedback or clear attitudes on items such as ratings, likes, and dislikes (example datasets are MovieLens (Cantador et al. 2011), Amazon (McAuley et al. 2015; He and McAuley 2016), and Douban (Ma et al. 2011a; Zheng et al. 2017)). Such connections most commonly appear in the form of a user–item matrix, where each row encodes the preferences of a user with his/her interacted items. Each element in the user–item matrix represents the user interaction with the item, which can be binary numbers if implicit interactive data were found or non-negative numbers if explicit interactive data were found.

**Contextual Data in GLRSs.** Context refers to the information collected during the interaction between the user and the item, such as timestamps, locations, or textual reviews, serving as an additional information source appended to the user–item interactions. Contextual data[8] are rich information attached to an individual user or item that depicts user characteristics such as job and gender (in e.g. dataset) or item properties such as description and product categories (in e.g. Foursquare (Gao et al. 2012), Yahoo! traffic stream (Menon et al. 2011) datasets). Such information associated with the user/item can form natural connections in certain relationships and therefore result in a network structure that can be used as GLRSs. In addition to the directly collected data attached to users/items, indirect information can also be obtained by a preliminary analysis of the dataset, such as user similarities, and entities derived from texts (Phuong et al. 2019). Undoubtedly, there exist obvious relationships between the implicitly derived information and the analysed user/item. For instance, the user similarity exhibits a hidden connection between users, and the entities extracted from an item text are originally attached to the item. It is indisputable that by rationally taking the graphs formed from contextual data, it can enrich information sources, provide more possibilities for graph learning-based recommendations, and improve the performance other than the accuracy of GLRSs (e.g. diversity).

**External Data in GLRSs.** To obtain more valuable resources apart from user–item interactions and their attached data from datasets, one can also seek external sources for GLRSs. A typical example can be found in knowledge graphs (KGs) which are graph structured data that describe entities or concepts and connect them

[8] Other terms may be used to indicate contextual data interchangeably such as side information, features, demographic data, categories, contexture information, etc. (Chen et al. 2020b). We do not distinguish them in this paper due to the same mathematical representation they share.

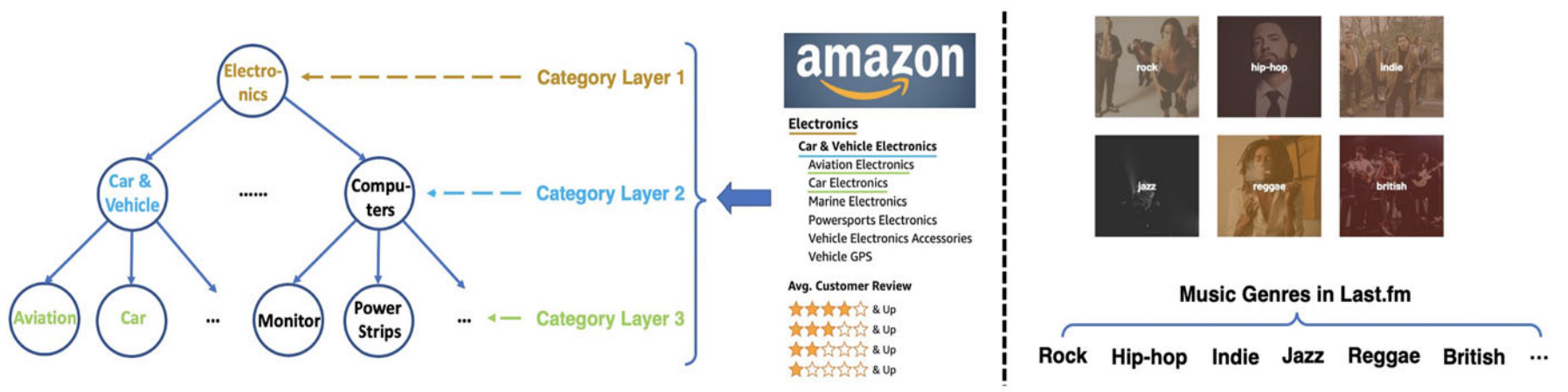

**Fig. 2** An example tree structure from the Amazon and Last.fm websites. From the root node to the leaf node, the hierarchical structure gradually refines and embodies the types of goods and music

with different types of semantic relationships (Liu et al. 2019a). External data can also provide complementary information to overcome data sparsity issues though some of them e.g. KGs may require domain knowledge for recommendation in specific domains such as e-commerce. Cross-domain knowledge is another type of external data, which refers to user/item side information from multiple sources. For instance, user profiles across different networks connected through anchor links (e.g. the link which connects the same entity from different platforms is called an anchor link) (Wang et al. 2017), item profiles from different communities (Farseev et al. 2017), user social relationships through information sharing platforms (Shi et al. 2021) are leveraged to improve both recommendation accuracy and diversify recommendation output. Accordingly, multiple subgraphs are built upon various sources which are then jointly learned for recommendation tasks (Wang et al. 2017).

Depending on different data shown above, one or more specific graph structures can be applied to form the input network to GLRSs. Based on this, we further innovatively propose to classify different graph types as tree-based graphs, homogeneous graphs, K-partite graphs, complex heterogeneous graphs, hypergraphs, and multiple graphs. The relations between different data and graph types are shown in Table 5 ("Appendix A").

### 4.1 Tree-based graphs

A tree-based graph where the items are organized in a hierarchy by a certain attribute of them (e.g. the category) is a natural yet powerful structure for human knowledge. It provides a machine- and human-readable description of a set of items and their parallel or hierarchical relationships like *affiliatedTo*, *subClass* and *isAPartOf* relations. Such hierarchical relations between items have been widely studied and proven to be effective in generating high-quality recommendations (Menon et al. 2011; Koenigstein et al. 2011; Mnih 2012; Kanagal et al. 2012; He et al. 2016; Yang et al. 2016; Sun et al. 2017). Tree-structured data are mostly obtained in a user/item context in **explicit contextual data**. The most common example is the categories of items. Typical domains of tree-based graphs in GLRSs consist of online products [e.g. the Amazon web store (McAuley et al. 2015)], foods (e.g. Gowalla (Liu et al. 2013)), movies (e.g. IMDB), and music (e.g. Last.fm).

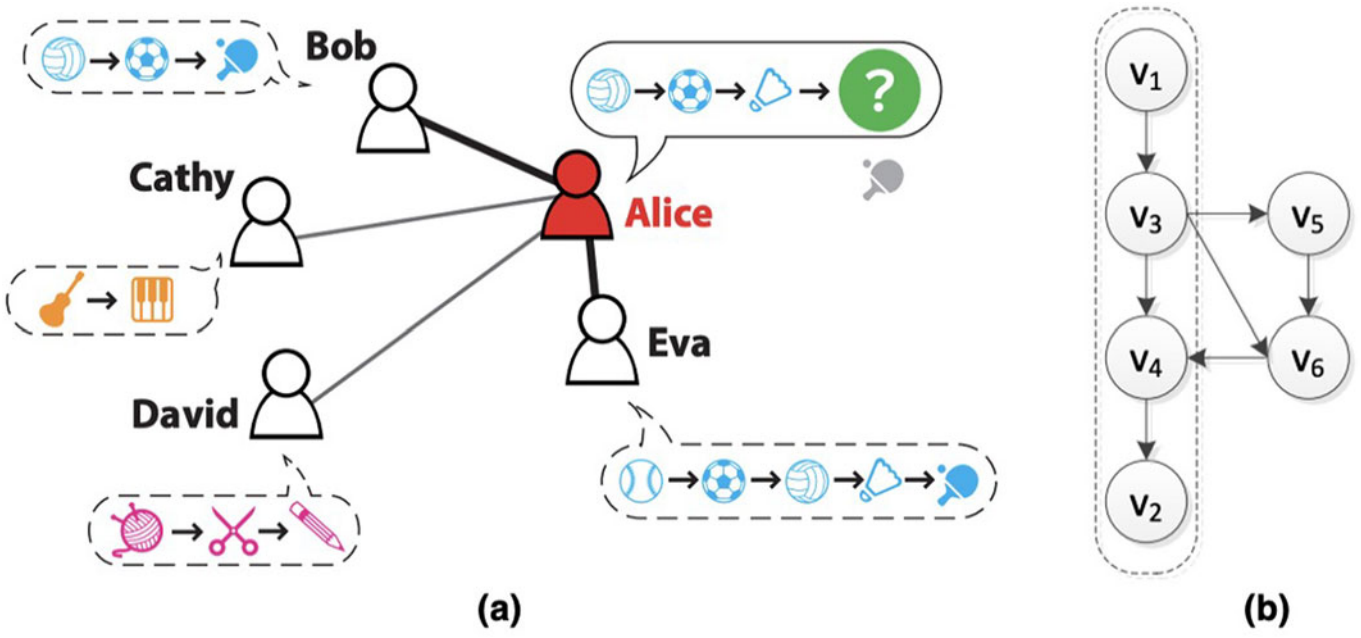

**Fig. 3** Homogeneous graph illustration. **a** An example attributed homogeneous graph with nodes representing users and edges representing social relations between users in Song et al. (2019b). Node attributes are from implicit interactive data. **b** An example of a session graph with nodes representing interacted items and edges connecting consecutive interacted items in one session (Xu et al. 2019a)

Figure 2 illustrates an example of tree-based graphs in Amazon and Last.fm to organize electronics or music by categories/genres. If a user buys a Monitor, she may possibly prefer Power Strips to match her Monitor instead of Aviation Electronics. This is due to both Monitor and Power Strips belonging to a higher layer category—Computers according to their intrinsic electrical characteristics. If a user prefers one song under a certain genre, she is more likely to favor other songs under this genre.

Representative algorithms coping with such graphs include Latent Factor Models (LFM) (Kanagal et al. 2012), Graph Distributed Representation-based Techniques (GDRM) (He et al. 2016), Deep Neural Networks (DNN) (Huang et al. 2019), and Attention Mechanism (AM) (Gao et al. 2019b).

## 4.2 Homogeneous graphs

A homogeneous graph is a graph with a single type of objects and links in GLRSs. Typical examples are user graphs in social networks, which include different types of social relations between users (Verma et al. 2019; Fan et al. 2019b). User–user relations can be derived from **interactive data** if recommending other users as the ultimate goal of the recommender system such as friend recommendation (Fan et al. 2019b), or **contextual data** if social relations are potential critical factors affecting users' next choices (Fan et al. 2019b; Farseev et al. 2017). The hypothesis behind taking a user social effect as essential contextual information is that two connected users in a user graph usually share similar preferences and influence each other by recommending items. Besides, various relationships among items, e.g. item co-occurrence and substitution items (Wang et al. 2020b; Xin et al. 2019; Gori et al. 2007) in a user behaviour sequence, connect all the items together and thus result in a homogeneous item graph. The co-occurrence relations in item graphs implicitly derived from **interactive data** not only reflect certain latent relations between items, but also reveal some behaviour patterns of users. It has been proved that the fusion of co-occurrence relations between items can yield significant performance enhancements (Wang et al. 2020b). With the development of Graph Neural Network (GNN) and its varieties,

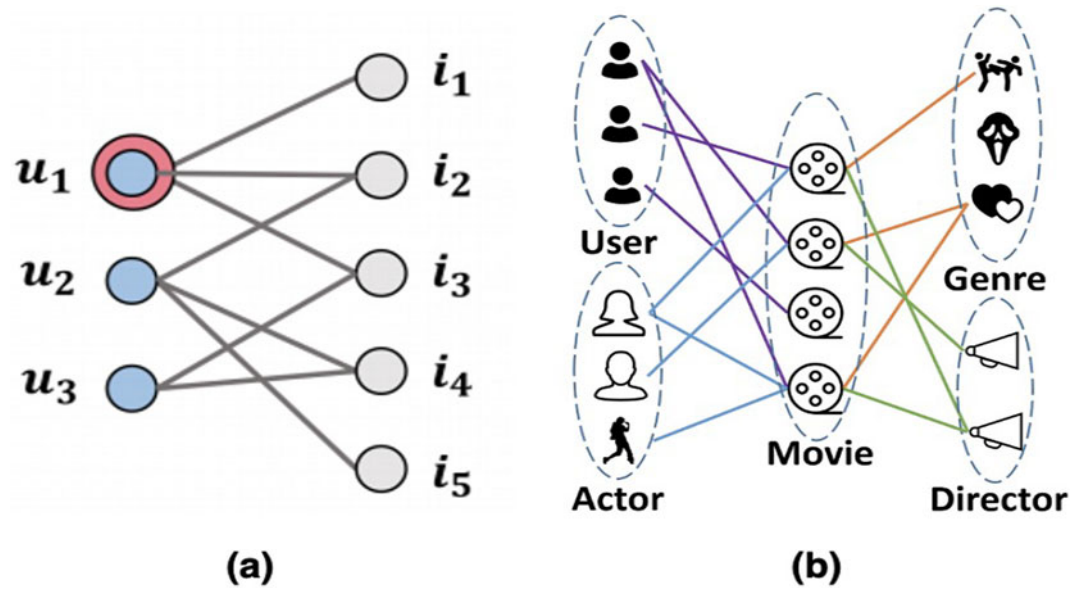

**Fig. 4** K-partite graph illustration. **a** A user–item bipartite graph with nodes representing user and items and edges representing interactions between users and items in Wang et al. (2019f). **b** A k-partite graph with nodes separated into four parts representing user, genre, movie, actor and director, and edges only existing among different node sets (Jiang et al. 2018a)

it becomes possible to capture consecutive transitions among nodes while generating accurate node embeddings for recommendation use. Based on this, in Xu et al. (2019a), Wu et al. (2019c), Abugabah et al. (2020), by considering the dynamic transitions in the interactive sequence within one session, the authors construct directed homogeneous session graphs in which a node represents an interacted item and an edge connects adjacent interacted items while retaining the order of interaction. The nodes of these homogeneous graphs are without attribute information, which is referred to as *non-attributed homogeneous graphs*. In practice, many real-world networks usually have attributes with their nodes that are also important for making sense of modelling network topological as well as contextual information for recommendation purposes. Such networks with node attributes and a single type of nodes as well as edges are named *attributed homogeneous graphs*[9] (Gao et al. 2018; Jamali and Ester 2009). An example can be found in a friend network where edges represent friendship (e.g. follow, like) between two users, and nodes represent users with attributes e.g. demographic information, or a sequence of items the user interacted with (Song et al. 2019b). In such a case, both social influence and user attributes can help to learn user preferences and thus affect the recommendation performance (Song et al. 2019b). Figure 3 illustrates examples of non-/attributed homogeneous graphs to help resolve the cold start issue of recommender systems.

Typical approaches in coping with such graphs include PageRank-based Models (PRM) (Gori et al. 2007), Translation-based Embedding Models (TEM) (Wang et al. 2020b; Gao et al. 2018), Random Walk-based Models (RWM) (Jamali and Ester 2009), Deep Neural Networks (DNN) (Xu et al. 2019a; Isufi et al. 2021), Graph Neural Networks (GNN) (Qiu et al. 2020b), and Deep Hybrid Models (DHM) (Wu et al. 2019c; Abugabah et al. 2020; Song et al. 2019b; Chen et al. 2019a).

### 4.3 K-partite graphs

A *k*-partite graph, also called multipartite graph, is a kind of graph whose nodes can be partitioned into *k* different independent sets so that no two nodes within the same set are adjacent. In the GLRSs scenario, the *k*-partite graphs are *unipartite graphs* when

[9] Some articles also categorize graphs into directed and undirected graphs. In our point of view, the undirected graph can be readily converted into a directed graph by replacing each edge with two oppositely directed edges. Thus, in this survey, without loss of generality, we assume that all graphs are directed graphs.

$k = 1$, *bipartite graphs* when $k = 2$ and *tripartite graphs* when $k = 3$. In Xin et al. (2019), the authors constructed unipartite graphs where nodes are items, and different relations exist between two nodes. Relations can be extracted from **contextual data** or **external sources** like knowledge graphs. Especially, bipartite graphs have attracted significant attention in areas like social network analysis (Tay and Lin 2014). They divide network nodes into two types and edges exist only between different types of nodes. User–item interactions can be naturally considered a bipartite graph, where the nodes represent users and items, and user nodes are linked with those interacted item nodes. Figure 4a gives an example of a bipartite graph formed from user–item interactive data (Wang et al. 2019f). The edges of the bipartite graph can either be a single type or multiple types of interactions, e.g. click, like, purchase or view (Li and Chen 2013; Zheng et al. 2018; Wang et al. 2019f; Zhang et al. 2019a; Phuong et al. 2019; Nikolakopoulos and Karypis 2019; Gori et al. 2007; Sun et al. 2020). In addition to user–item interactions, auxiliary information of user/item can also be constructed as a bipartite graph. For instance, items and their attributes (e.g. pin-boards for Pinterest dataset (Ying et al. 2018)) can also be seen as two types of nodes in forming an item–entity bipartite graph.

Apart from user–item interactions, bipartite graphs can be extended to multi-partite graphs by mining the contextual information of user/item and leveraging beneficial attributes or related information such as user query, actors and actresses of movies as other groups of nodes (Kim et al. 2019; Fan et al. 2019a; Berg et al. 2017; Cheng et al. 2007; Lei et al. 2020a; Jiang et al. 2018a). For instance, the authors of Cheng et al. (2007) built a user–movie–genre tripartite graph for personalized recommendation. In Fig. 4b, a multi-partite graph is formed by utilizing contextual information of items (movies), such as genre, actor, director (Jiang et al. 2018a).

$K$-partite graphs can be weighted by assigning numerical points like a rating or similarity score on corresponding edges (Chen et al. 2020a; Zhang et al. 2019a; Yao et al. 2015). In Yao et al. (2015), the authors constructed a multi-partite graph incorporating user, item, and their related contextual features. Each number on the edge represents the co-occurrence of two end-nodes. Figure 4 shows examples of k-partite graphs in the GLRS scenario.

Various GLRS-based approaches have been proposed to learn such kind of graphs, such as Kernel-based Machine learning methods (KML) (Li and Chen 2013), RWM (Eksombatchai et al. 2018; Jiang et al. 2018a), Meta-path based Methods (MPM) (Yu et al. 2014; Lu et al. 2020; Yu et al. 2013), GDRM (Cen et al. 2019), TEM (He et al. 2015), DNN (Zheng et al. 2018), Auto-Encoder (AE) (Zhang et al. 2019a), AM (Wang et al. 2019d), GNN (Wang et al. 2019f; Chen et al. 2020a) (see Table 3 for more details).

### 4.4 Complex heterogeneous graphs

Complex heterogeneous graphs are related to graphs with multiple types of nodes and edges. The connections among nodes do not follow specific rules (Xu et al. 2019b; Jiang et al. 2018b; Shi et al. 2018; Lu et al. 2020; Wang et al. 2020f; Kyriakidi et al. 2020; Feng and Wang 2012; Zheng et al. 2017). For instance, in Wang et al. (2020f)

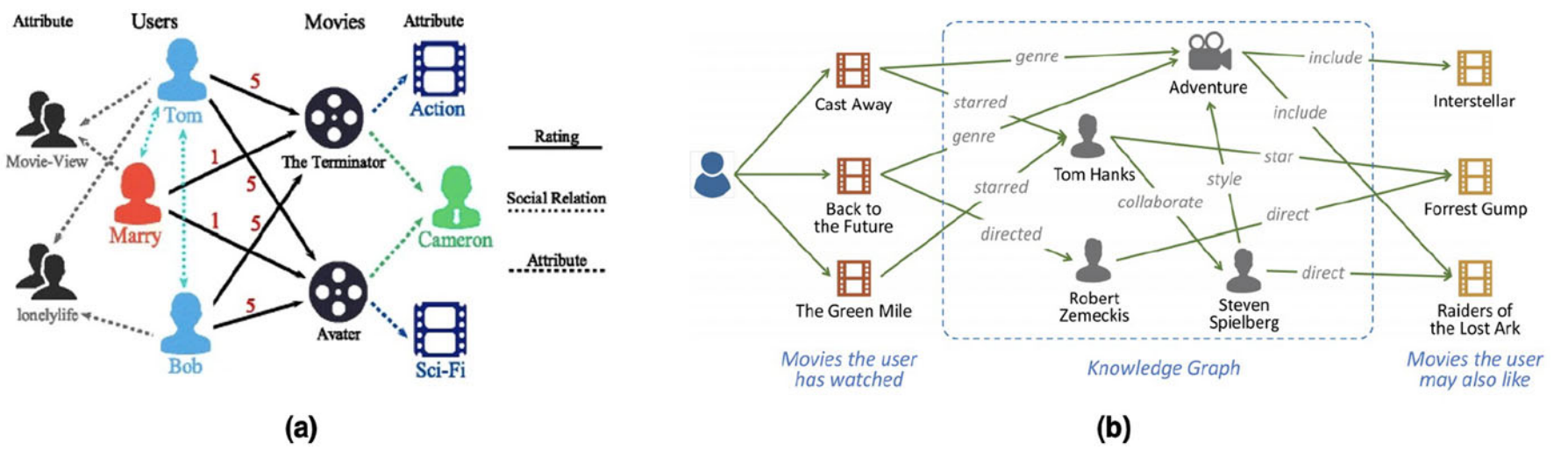

**Fig. 5** Complex heterogeneous graph illustration. **a** An example of complex heterogeneous graphs in which there are three kinds of nodes representing group, user, and item, respectively, and group–user, user–item two kinds of edges, for social recommendations (Shi et al. 2018). **b** An illustration of KGs with multiple kinds of entity nodes relation edges for exploring high-order user preferences for recommendations (Wang et al. 2019a)

the authors constructed long- and short-term graphs in which nodes are divided into user and item nodes, and edges can exist either between user and item, or among items. Complex heterogeneous graphs are usually non-weighted by default, but they can also be weighted. In Zheng et al. (2017), Shi et al. (2016), numbers on user–item edges depict the user's rating on the item. Figure 5a, b gives examples of weighted and unweighted complex heterogeneous graphs, respectively.

The goal of complex heterogeneous graph learning is to obtain the latent vertex representations by mapping vertexes into a low-dimensional space, which then can be leveraged for recommendations (Xu et al. 2019b; Jiang et al. 2018b; Shi et al. 2018; Lu et al. 2020). The authors of Jiang et al. (2018b) constructed a complex heterogeneous graph with four types of nodes and ten types of edges. The node representations are learned for citation recommendations. In Shi et al. (2018), the authors constructed a heterogeneous graph with user, item, and various types of attributes as nodes. Finally, the learned user/item representation is adopted to predict the user's rating score of the item to make recommendations. Differently, in Kyriakidi et al. (2020), the connection mode can be revealed via path traversing, so that the similarities between two nodes can be calculated for a recommendation.

Another typical example falling into this category is a **Knowledge Graph (KG)**. KG is a multi-relational graph composed of entities as nodes and relations as different types of edges as illustrated in Fig. 5b. Each edge of KG represents a triple of the form *(head entity, relation, tail entity)*, also called a *fact*, indicating that two entities are connected by a specific relation. Recent years have witnessed a rapid growth in KG application in recommendation, resulting in promising improvements in both recommendation accuracy and explainability due to the rich structured information that KG provides about the items. Existing KGs, e.g. Yago (Suchanek et al. 2007), DBPedia (Lehmann et al. 2015), provide auxiliary information apart from user–item interactions. The relational properties in KGs break down the independent interaction assumption by linking items with their attributes. Meanwhile, the introduction of KGs alleviates the data sparsity and cold-start issues raised in recommender systems (Wang et al. 2019b; Ma et al. 2019; Wang et al. 2020g; Zhou et al. 2020b; Song et al. 2019a; Yang and Dong 2020; Wang et al. 2020a; Ai et al. 2018; Zhang et al. 2016; Ma et al. 2019; Catherine and Cohen 2016; Shi et al. 2020; Chen et al. 2019d; Wang et al. 2020g;

Zhou et al. 2020a, b; Cao et al. 2019; Yu et al. 2013; Ostuni et al. 2013). However, various types of entities and relations in KGs also pose the challenge of capturing semantically interconnected information for effective recommendations. In addition, how to reasonably and vividly provide recommendation results through KG internal reasoning and the linkage among user–item interactions deserves more attention (Wang et al. 2019g), for instance, how to impartially and convincingly explain the reasoning process of the recommendation list to the target user. The resources of all KGs used for GLRSs have been collected and are displayed in Table 4.

Typical technologies related to learn complex heterogeneous graphs include LFM approaches (Zhao et al. 2017), PRM (Catherine et al. 2017; Catherine and Cohen 2016), RWM approaches (Bagci and Karagoz 2016; Ma et al. 2019), Rule-based Model (RM) (Shi et al. 2020), MPM (Shi et al. 2015; Ostuni et al. 2013), GDRM (Shi et al. 2018; Fu et al. 2020; Palumbo et al. 2017; Jiang et al. 2018b), TEM (Chen et al. 2019d; Wang et al. 2018b, 2019a; Ai et al. 2018), AE (Zhang et al. 2016), AM (Han et al. 2018), Graph Neural Network (GNN) (Wang et al. 2019e, c), Deep Reinforcement Learning (DRL) (Song et al. 2019a; Lei et al. 2020a), and DHM (Sheu and Li 2020; Wang et al. 2020g; Zhou et al. 2020a; Wang et al. 2018c). Other techniques such as Spectral Clustering (Farseev et al. 2017) are also adopted to learn such kind of graphs. (Refer to Table 3 for more details)

### 4.5 Hypergraphs

Hypergraphs are defined as a generalization of graphs in which the edges are arbitrary non-empty subsets of the vertex set (Agarwal et al. 2006). Instead of having edges between pairs of vertices, hypergraphs have edges that connect sets of two or more vertices. Correspondingly, such edges of hypergraphs are called *hyperedge*. If the hyperedge always has the degree of 2, the hypergraph reduces to an ordinary graph. For clarity, we only refer to hypergraphs with multiple hyperedge degrees. The motivation for introducing hypergraphs for recommendations is twofold (Feng et al. 2019): first, the data correlations can be more complex than the pair-wise relationship, which is difficult to model with traditional graph structures; second, the data representations can be multi-modal, which means that data can be connected through e.g. text information, visual information, or social connections, which is difficult to capture with the traditional graphs. Thus, a hypergraph is a way to model a more general data structure, and the recommendation performance can be improved through the modelling of the high-order proximity in the constructed hypergraphs (Yu et al. 2021).

Recent years witness the learning of hypergraphs in promoting the development of recommender systems (Wang et al. 2020c; Bu et al. 2010; Li and Li 2013; Tan et al. 2011; Zhu et al. 2016; Mao et al. 2019; Yu et al. 2021; Gharahighehi et al. 2021, 2020). Hypergraphs can be generalized explicitly from data sources or derived implicitly indirectly through, for instance, a clustering technique. In Bu et al. (2010), the authors constructed a hypergraph with six types of vertices and nine types of hyperedges representing complex relationships of nodes rather than a pair-wise form. Specifically, user–item **interactive data**, **contextual data** such as social relations, tagging and album information, are adopted. Meanwhile, the group information obtained through

**Table 4** A collection of commonly used knowledge graphs

| | Name | Domain Type | Main Knowledge Source |
|---|---|---|---|
| **General KG** | YAGO (Suchanek et al. 2007) | Cross-Domain | Wikipedia |
| | Freebase (Bollacker et al. 2008) | Cross-Domain | Wikipedia, NNDB, FMD, MusicBrainz |
| | DBpedia (Lehmann et al. 2015) | Cross-Domain | Wikipedia |
| | Satori[20] | Cross-Domain | Web Data |
| | CN-DBPedia (Xu et al. 2017) | Cross-Domain | Baidu Baike, Hudong Baike, Wikipedia (Chinese) |
| | NELL (Carlson et al. 2010) | Cross-Domain | Web Data |
| | Wikidata[21] | Cross-Domain | Wikipedia, Freebase |
| | Google Knowledge Graph[22] | Cross-Domain | Web data |
| | Facebooks Entities Graph[23] | Cross-Domain | Wikipedia, Facebook data |
| | ConceptNet (Speer et al. 2017) | Cross-Domain | Web data |
| | MultiWordNet (Pianta et al. 2002) | Cross-Domain | Princeton WordNet |
| | Babelfy (Moro et al. 2014) | Cross-Domain | BabelNet |
| | Open Multilingual Wordnet (Bond and Paik 2012) | Cross-Domain | Wiktionary, Unicode Common Locale Data Repository |
| **Domain Specific KG** | Bio2RDF (Belleau et al. 2008) | Biological Domain | Public bioinformatics databases, NCBIs databases |
| | KnowLife (Ernst et al. 2014) | Biomedical Domain | Scientific literature, Web portals |
| | IMDB[24] | Movie Domain | Web data |
| | KnowIME (Yan et al. 2020) | Intelligent Manufacturing Domain | Internet, Baidu Encyclopedia, and related intelligent manufacturing websites |

K nearest neighbour and K-means also forms part of the hyperedges. Pliakos and Kotropoulos (2014) used a unified hypergraph model boosted by group sparsity optimization and encapsulated the high order connections among users, images, tags, and geo-information for tag recommendation. An hyperedge can be weighted by binary values indicating whether to participate or whether the link exists between two nodes, or integer values indicating the frequency of participation (Yu et al. 2021). Besides,

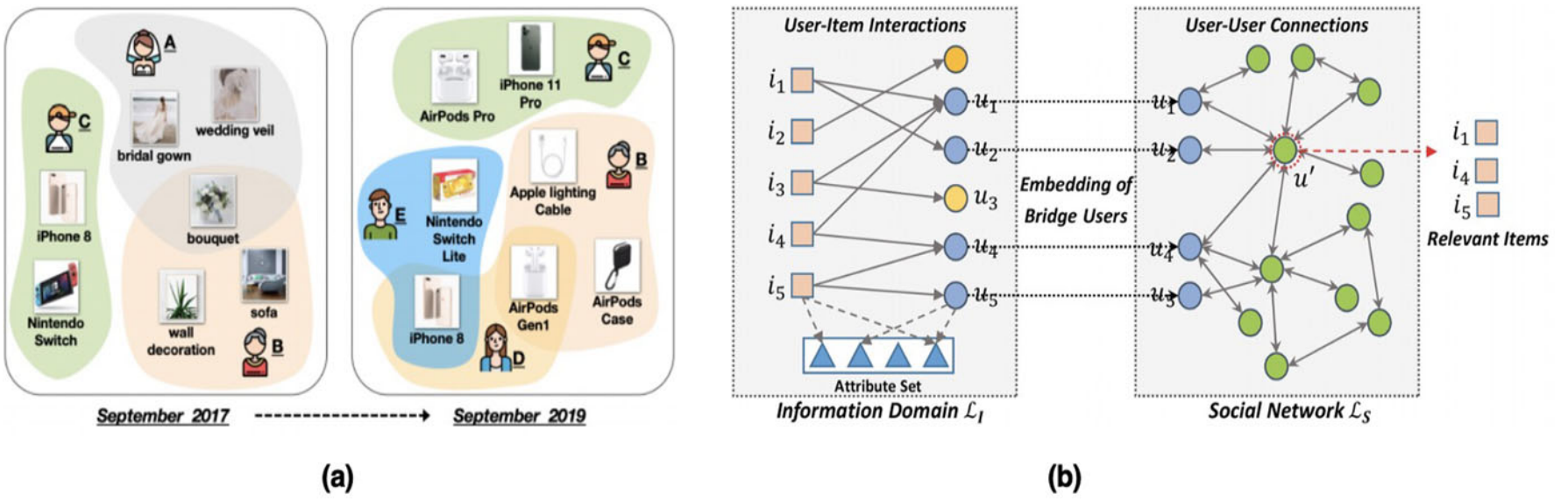

**Fig. 6** **a** An example hypergraph in which nodes are represented in different colours and edges exist between colours of nodes (Wang et al. 2020c). **b** An example of multiple graphs in which there is a user–item–attribute tripartite graph and a homogeneous social graph (Wang et al. 2017)

decimal values such as similarity scores can also be assigned as edge weight depending on recommendation context (Bu et al. 2010; Tan et al. 2011). Different weights assigned to a hypergraph differentiate the impact of each hyperedge. An example of a hypergraph is illustrated in Fig. 6a, in which the hypergraph consists of three nodes with one hyperedge in September 2017 and four nodes with two hyperedges in September 2019 (Wang et al. 2020c).

Representative approaches in dealing with hypergraphs in GLRS include Regularization Theory-based Graph Ranking (RTGR) (Bu et al. 2010), GNN (Yu et al. 2021), and DHM (Wang et al. 2020c; Gharahighehi et al. 2020). More specific details are found in Sect. 5 and Table 3.

### 4.6 Multiple graphs

Incorporating various types of information involved into a graph will often make the composed graph too complicated and not conducive to subsequent algorithms. Therefore, many researchers choose to split complex heterogeneous graphs into multiple subgraphs for learning separately. While such a “divide and conquer" strategy disassembles the complexity of the graph, it can also be adapted to certain graph learning algorithms (Verma et al. 2019; Fan et al. 2019b; Wu et al. 2019b; Vijaikumar et al. 2019; Wang et al. 2017). For instance, in Fan et al. (2019b), the authors disassembled the complex multigraph with more than one type of edge between two nodes, into three homogeneous subgraphs, which are then learned separately through the DeepWalk technique for friend recommendation. More researches choose to separate **interactive data** from **contextual data/external data** as a powerful supplementary information for learning user/item representations and meanwhile to some extent alleviating the cold-start issue (Liu et al. 2020; Ali et al. 2020). The authors of (Fan et al. 2019b; Wu et al. 2019b) extract the user social relations from original data to build a homogeneous graph of the social network. Similarly, in Fig. 6b, the authors of Wang et al. (2017) use the users as anchor links (e.g. the link which connects the same entity from different platforms is called an anchor link) to obtain the user social information from cross-domain knowledge resulting in homogeneous subgraphs to improve recommendation performance. In Monti et al. (2017), Isufi et al. (2021), the authors construct two homogeneous graphs, namely an item graph and a user graph in which nodes represent items and users, respectively, and edges represent similarities between adjacent nodes.

Another advantage of leveraging a multi-graph for recommendation is scalability. When a new data source is added, it can be constructed as a separate graph independent of the original data source, and the learned embedding representations can be merged in the upper layer. In such cases, different data sources can be separated into a more concise and clear form for subsequent maintenance and learning (Vijaikumar et al. 2019).

Various ways can be adopted to learn multiple graphs for recommendations. For graphs with similar structures (e.g. either homogeneous or heterogeneous), similar learning strategies are usually adopted to obtain the low-dimensional representation of nodes (Ali et al. 2020; Monti et al. 2017), while it is usually necessary to adopt different learning strategies for graphs with different structures. This is reasonable because usually a certain learning strategy should be performed under certain assumptions or rules (Ma et al. 2011a; Chen et al. 2019a).

To deal with multiple graphs, the most common way is to learn each independent graph separately and then aggregate the results for recommendation purposes. Typical examples are GDRM (Verma et al. 2019; Ali et al. 2020), LFM (Ma et al. 2011a), AM (Vijaikumar et al. 2019), GNN (Fan et al. 2019b; Wu et al. 2019b), DHM (Monti et al. 2017), and DNN (Wang et al. 2017).

### 4.7 Graph discussion

Graphs, ranging from a flat tree-based structure to a complex network structure, from a homogeneous network to a heterogeneous one, evolve from both structural and contextual sides. As summarized in "Appendix Table 5", though many different datasets overlap across various graphs and recommendation tasks, it is undeniable that there is no perfect graph type that can embrace all types of data or solve all problems of the recommender system. However, we can still make several observations:

First, tree-based graphs are mainly derived from e-commerce datasets (e.g. Amazon and JD) in which types of commodities can be broken down by the category's level of granularity. Such kind of tree structure provides recommender systems with a paradigm which can be further refined in a "top-down" manner.

Second, homogeneous graphs can be formed by leveraging user social networks (user graph) or interlinking of items (item graph). A session graph is a kind of directed item graph with nodes of items that are clicked by the user and links of relations that are according to the clicked order. Homogeneous graphs can be used to mine the relationships between a single type of node (e.g. users/items/sessions) to make targeted modelling for recommendations.

Third, most k-partite graphs, especially for $k = 2$, are constructed based on user–item interactions, which are the preliminary requirements for RS and can be naturally formed into a graph with users and items as two groups of nodes. For generalized k-partite graphs, they normally formalize by extending the user–item bipartite graph with contextual knowledge w.r.t. users/items that are deemed to potentially benefit recommendation performance.

Besides, complex heterogeneous graphs incorporating multi-types of nodes and relations reveal the intricate relationship between user, item, and other related information from contextual/external data sources in real scenarios. Especially, KGs are generally leveraged in the domains of movies, music, books, or news, where named entities are highly recognizable and have been witnessed, and most of which can be found in the corresponding entries in e.g. Wikipedia, DBpedia, Yago, etc.

Furthermore, hypergraphs have recently been introduced to represent the connections between sample groups such as user groups according to social relationships, or item groups according to a user's co-purchase history, which breaks the traditional node-to-node pattern, pursuing a higher-level representation of data structure. With their diversity, heterogeneous graphs with various aspects of information can be used to model more complex situations than other types of graphs so as to better solve the cold-start and data sparsity issues of recommender systems. However, such complexity of both graph structure and content leads to a more complicated modelling process and brings challenges that cannot be ignored. One possible solution is to break down such complexity into multiple subgraphs. In this way, the complex relationship of the graph is degraded, and the relationship between nodes is more concise and clear. To reveal the group knowledge among nodes and meanwhile incorporate multi-modal information for better recommendations, a hypergraph is proposed to express multiple relationships beyond the pair-wise relations. With the development of technology, the types of graphs that computers can process tend to be increasingly complex and fine-grained, which suggests that RS can handle more sophisticated and multi-dimensional problems and scenarios.

## 5 Graph-based models for GLRS

In this section, we will go through various models adopted in GLRS and analyse them from the technical perspective, as shown in Fig. 7. Many GLRS recommendation algorithms are not only based on a specific technology but often include multiple different types of technologies. Therefore, our classification is mainly based on the technology used by the key components of the model pointed out in the paper. A comprehensive technology summary in GLRS is found in Table 6 ("Appendix B").

### 5.1 Traditional techniques

Traditional machine learning techniques used in GLRS can generally be classified into PageRank-based models (PRM), regularization theory-based graph ranking techniques (RTGR), kernel-based machine learning techniques (KML), latent factor models (LFM), and others (e.g. association mining and clustering). Graph learning-based traditional techniques can be used for multiple types of graphs from various domains of data sources. The classic PageRank algorithm computes an importance score for every node according to the graph connectivity, which is usually exploited to tackle the recommendation as a ranking problem. Many adapted PageRank algorithms have been proposed and widely applied to a variety of graphs like homogeneous item graphs (Gori et al. 2007), K-partite graphs (Shams and Haratizadeh 2017; Jäschke et al. 2007; Musto et al. 2017, 2021), and complex heterogeneous graphs (Catherine and Cohen 2016; Lee et al. 2013) for recommendation. A separate line of work in this area—regularization theory-based graph ranking techniques—is widely exploited to learn the ranking results on hypergraphs for recommendations (Bu et al. 2010; Li and Li 2013; Tan et al. 2011; Mao et al. 2019). They usually use hypergraphs to model high-order relations among various types of objects in social networks and use the regularization framework (Zhou and Schölkopf 2004) for ranking graph data. Some research works use kernel-based machine learning techniques to compute similarities between structured objects, such as nodes of a graph, that cannot be naturally represented by a simple set of numbers, and demonstrate their effectiveness in a variety of graphs in RS (Li and Chen 2013; Yajima 2006; Li et al. 2014; Fouss et al. 2012; Ostuni et al. 2014) (homogenous graph, K-partite graph and knowledge graph). For instance, Yajima (2006) used a Laplacian kernel to capture the positional relations among nodes on a homogenous item graph and built one-class SVM models for each user to recommend items that are positionally closer to their previously bought items. Li and Chen (2013) propose a generic kernel-based machine learning approach of link prediction in bipartite graphs and apply it in recommender systems. In a later work, Li et al. (2014) inspect a spectrum of social network theories to systematically model the multiple facets of a homogeneous user graph in social networks and infer user preferences. They design and select kernels corresponding to major social network theories and then adapt a non-linear multiple kernel learning technique to combine the multiple kernels for recommendation. Latent factor models are usually adopted by many researchers to learn the latent feature of users and items for recommendation purposes. In GLRS, LFM can be accompanied by regularization terms constraining the trust/distrust relations between users to integrate contextual information extracted from users' social networks (Du et al. 2011; Ma et al. 2009; Sun et al. 2015). Ma et al. (2008), Ma et al. (2011b) proposed a factor analysis based on probabilistic matrix factorization to alleviate the data sparsity and poor prediction accuracy problems in recommender systems by incorporating social contextual information, such as users' social trust networks and social tags. In addition, there are many other traditional machine learning techniques used to model the relations between the nodes in GLRS, such as clustering (Farseev et al. 2017; Song et al. 2011), association mining (Huang et al. 2004) and graphing complex numbers (Xie et al. 2015), and applied to K-partite graphs (Song et al. 2011), complex heterogeneous graphs (Huang et al. 2004; Xie et al. 2015), and multiple graphs (Farseev et al. 2017).

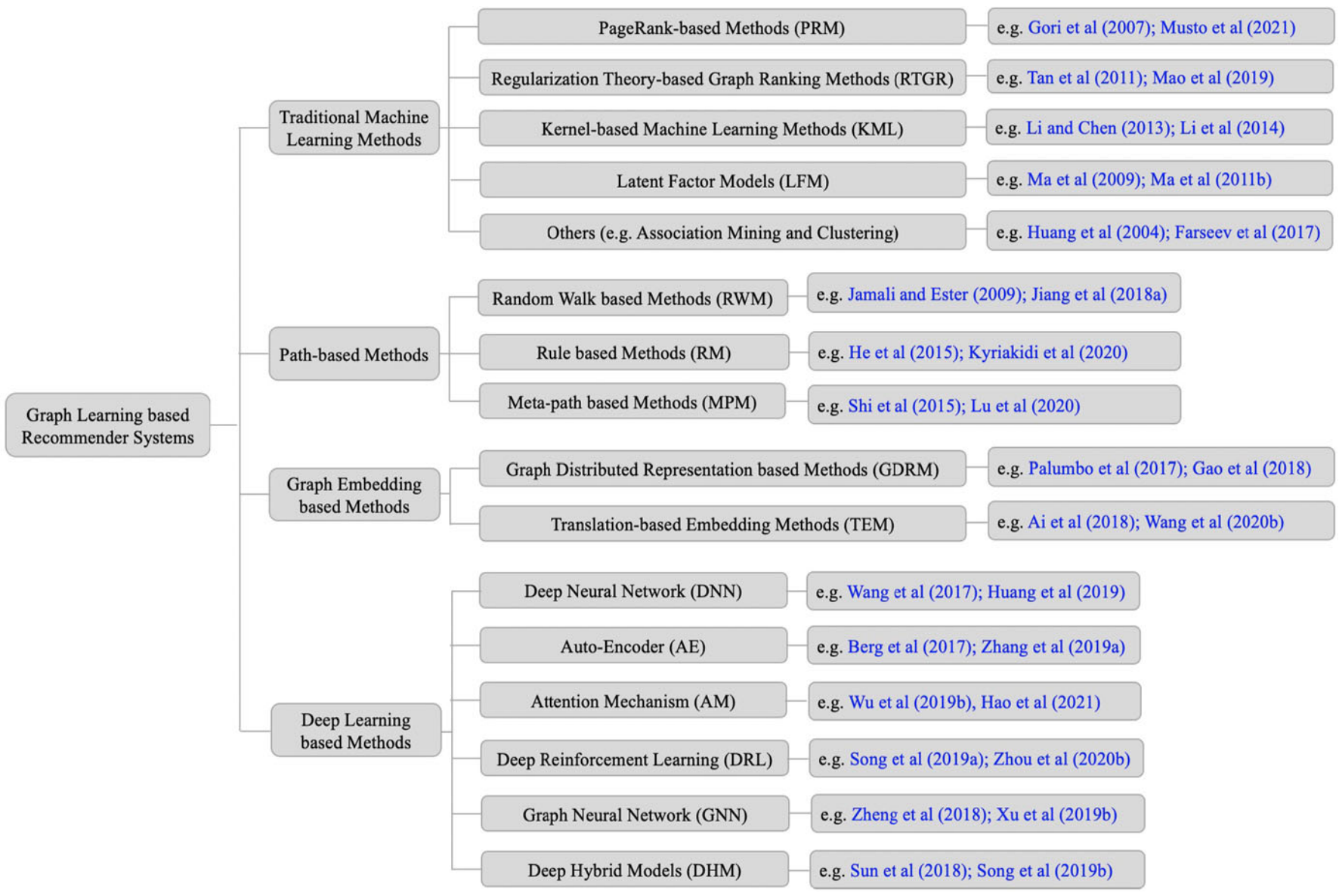

**Fig. 7** A categorization of GLRS techniques from the technical perspective

### 5.2 Path-based techniques

For heterogeneous graphs with multiple types of nodes and relations, the basic idea of earlier recommendation strategies is to leverage path-based semantic relatedness between users and items over the constructed graphs. Different from similarity-based techniques based upon the item/user attributes, path-based techniques especially emphasize the essential role of links in graphs, and links between start node and end node can form a path serving a recommendation purpose. In this case, the underlying relationships via network propagation show particularly important for indirectly connected objects. They can be used mainly for complex heterogeneous graphs (Bagci and Karagoz 2016; Ma et al. 2019; Feng and Wang 2012; Shi et al. 2015, 2018, 2020; Zheng et al. 2017; Shi et al. 2016; Ostuni et al. 2013) and k-partite graphs (Li and Chen 2013; Cheng et al. 2007; Yao et al. 2015; Jiang et al. 2018a; Nikolakopoulos and Karypis 2019; Sharma et al. 2016; Eksombatchai et al. 2018; He et al. 2015; Cen et al. 2019; Lu et al. 2020; Yu et al. 2013, 2014), but can also be adopted for multiple and homogeneous graphs (Yin et al. 2010; Vijaikumar et al. 2019; Jamali and Ester 2009; Gori et al. 2007), covering various domains such as POI recommendation, academic and book, e-commerce, and entertainment domains.

Earlier studies leverage a series of predefined rules to generate a path on the constructed graphs followed by different similarity measurements for ranking the candidate items for recommendation (He et al. 2015; Catherine and Cohen 2016; Catherine et al. 2017; Kyriakidi et al. 2020). Another graph tracing algorithm initially designed for homogeneous networks is a random walk-based algorithm (Andersen et al. 2008). It starts at a node and follows outgoing edges, uniformly at random or according to predefined transition probability, until the stop condition is reached. The output paths indicate the context of connected vertices. The randomness of walks gives the ability to explore the graph while considering both the global and local structural information by walking through neighbouring vertices. The random walk mechanism enables capturing complex, high-order, and indirect relations between nodes for recommendations. Due to these advantages, random walk and its various variants are favored for a long period in the GLRS domain for generating paths in homogeneous as well as heterogeneous graphs (Jamali and Ester 2009; Feng and Wang 2012; Jiang et al. 2018a; Eksombatchai et al. 2018; Nikolakopoulos and Karypis 2019; Gori et al. 2007; Li and Chen 2013; Yin et al. 2010; Vijaikumar et al. 2019; Sharma et al. 2016; Mao et al. 2019; Bagci and Karagoz 2016; Ma et al. 2019; Cheng et al. 2007; Yao et al. 2015). For instance, Jiang et al. (2018a) propose a generalized random walk with restart model on a k-partite graph to extract the paths. Then, a BPR-based machine learning technique is leveraged to learn the weights of links in the graph.

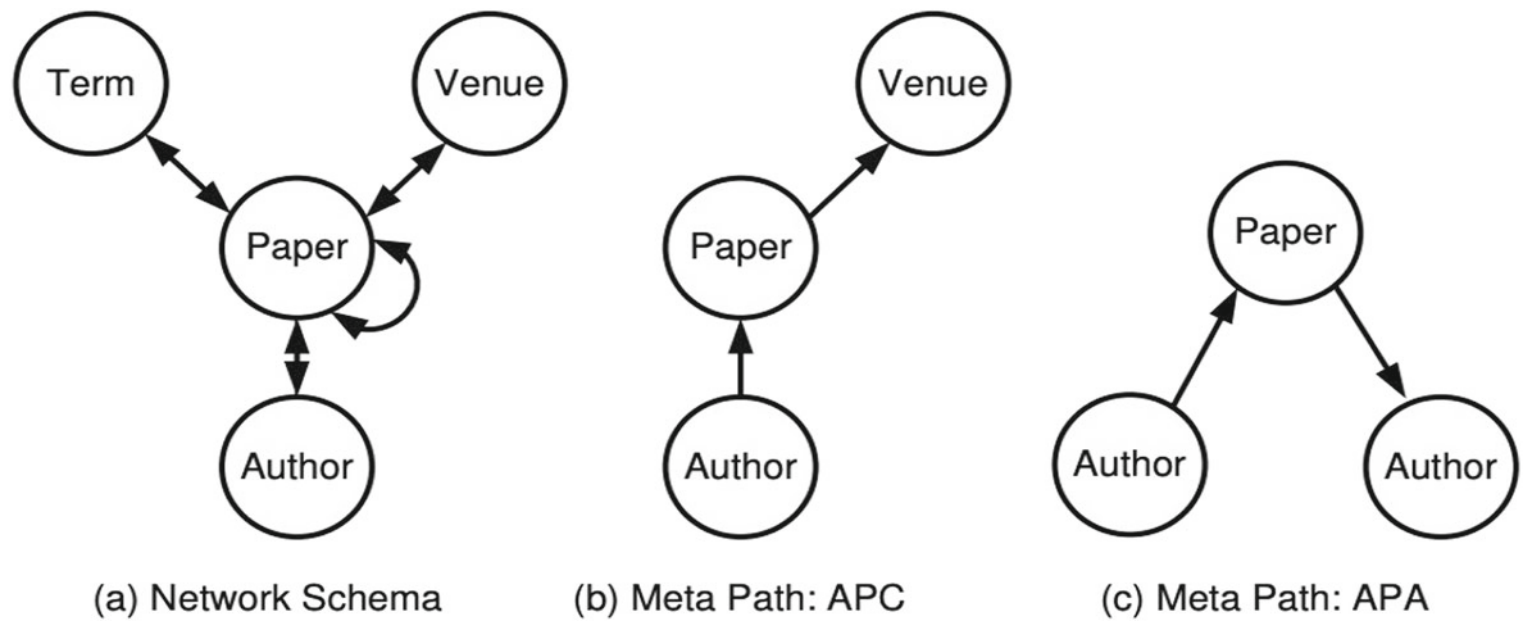

**Fig. 8** Bibliographic network schema and meta paths defined in Sun et al. (2011)

To integrate different types of objects and links in heterogeneous networks, the work of Sun et al. (2011) proposed the concept of a meta-path, which is adopted by many later researches (Yu et al. 2013; Cen et al. 2019; Lu et al. 2020; Shi et al. 2018, 2015; Zheng et al. 2017; Shi et al. 2016; Ostuni et al. 2013; Yu et al. 2014). Specifically, a meta-path is a path defined on the graph of network schema $\mathcal{T}_{\mathcal{G}} = (\mathcal{A}, \mathcal{R})$,[25] and normally denoted in the form of $A_1 \xrightarrow{R_1} A_2 \xrightarrow{R_2} A_3 \ldots \xrightarrow{R_l} A_{l+1}$, which defines a composite relation $R = R_1 \circ R_2 \circ \ldots \circ R_l$ between types $A_1$ and $A_{l+1}$, where $\rightarrow$ explicitly shows the direction of a relation from graph $\mathcal{G}$, $\circ$ denotes the composition operator on relations. Figure 8 illustrates two examples of meta-paths Fig. 8b, c derived from network schema Fig. 8a. When a user-specific meta-path e.g. $P = (A_1 A_2 \ldots A_l)$ has been given, several similarity measures can be defined for a pair-wise nodes comparison, namely to compare $v_i \in A_1$ and $v_j \in A_l$ according to a series of paths derived based on $P$, referred to path instances. Random walk is one representative to generate paths instances $p \in P$ following the predefined meta-path schema (Shi et al. 2015). To further learn the attributed heterogeneous information network (HIN) for better recommendations, later studies attempt to combine meta-paths with a traditional latent model, e.g. FM (Zhao et al. 2017), MF (Shi et al. 2018; Yu et al. 2014). Though random-walk-based similarity measures require less domain knowledge compared to meta-path-based measures,[26] the latter turn out to be more meaningful and interpretable in most GLRSs (Sun et al. 2011).

[25] Please refer to Definition 1 and 2 in Section 3.1 for the meaning of the symbols.

[26] Meta-path-based approaches usually require handcrafted features to represent path semantics and thus further require domain knowledge (Sun et al. 2011).

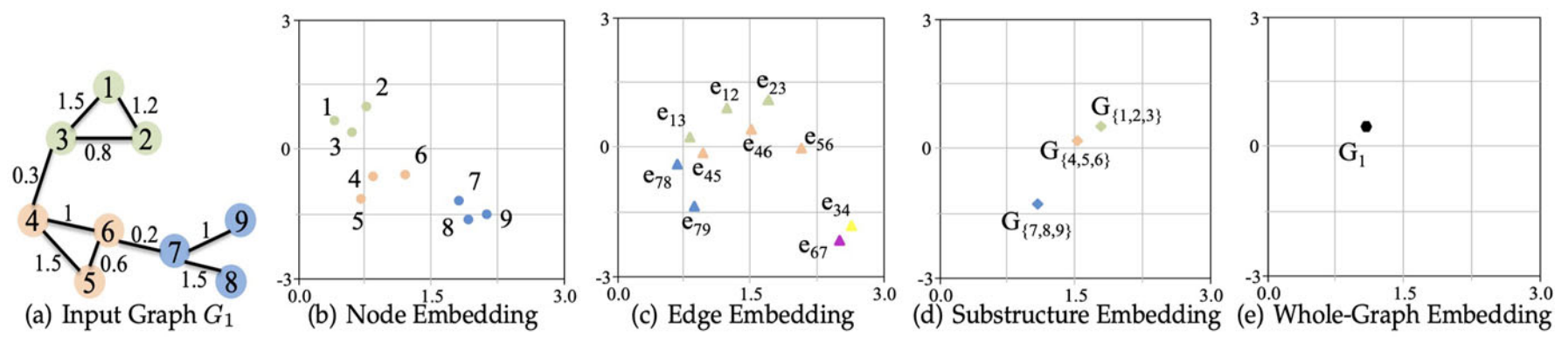

**Fig. 9** A toy example of embedding a graph into 2D space with different granularities (Cai et al. 2018). $G_{1,2,3}$ denotes the substructure containing node $v_1$, $v_2$, $v_3$

Despite that path-based similarity strategies have achieved initial success in improving RS accuracy to some extent, challenges still exist. First, meta-path-based similarities rely on explicit path reachability and the quality would be affected by the sparse and noisy input data, especially for links that are accidentally formed but do not convey meaningful information for recommendations. Second, the explicit path relatedness derived from the path-based similarity does not necessarily have a positive impact on recommendation performance. For instance, the work of Yu et al. (2013) learns a linear weighting mechanism to integrate the extracted meta-paths for the subsequent recommendations, ignoring the complicated mapping mechanism of the constructed k-partite graphs. Third, path-based similarity strategies need to generate similarity scores for all candidate items at each step for every user which reduces the effectiveness of the system and thus makes it difficult to be applied in a large-scale scenario.

### 5.3 Graph embedding-based techniques

The motivation for applying graph embedding (GE) strategies lies in that they can provide an effective yet efficient way to solve the graph analytics problem (Cai et al. 2018). GE-based techniques are mainly applied to complex heterogeneous graphs (Fu et al. 2020; Wang et al. 2020a; Jiang et al. 2018b; Palumbo et al. 2017; Wang et al. 2018b; Cao et al. 2019; Wang et al. 2021b; Cao et al. 2019; Ai et al. 2018; Chen et al. 2019d) from multiple data sources but can also be applied to homogeneous graphs (Gao et al. 2018; Wang et al. 2020b), tree graphs (He et al. 2016), k-partite graphs (Li et al. 2021b), and multiple graphs (Verma et al. 2019; Ali et al. 2020) in different recommendation domains.

Specifically, graph embedding converts a graph into a low dimensional space in which the graph information can be retained as much as possible. By representing a graph as a (or a set of) low-dimensional vector(s), graph algorithms can be applied efficiently. Figure 9 illustrates how graph embedding projects a graph into the vector space with different granularities, e.g. w.r.t. node/edge/substructure/whole graph (Cai et al. 2018). Some researches differentiate graph representation learning and graph embedding by comparing the dimension of the output embedding vectors with the dimension of the inputs (Cai et al. 2018). Graph embedding focuses on learning the low-dimensional representations, while graph representation learning does not require the learned representations to be low dimensional. Though they have slight differences, we do not make a special distinction in this survey. Essentially, the two approaches aim to project a graph into the vector space while preserving the graph structure and capturing the connectivity information within the graph to serve the recommendation task. The mapping can be defined as:

$$f : v_i \rightarrow \boldsymbol{x}_i \in R^d \tag{2}$$

where $d \ll |\mathcal{V}|$, and $\boldsymbol{x_i} = \{x_1, x_2, \ldots, x_d\}$ is the embedded or learned vector that captures the structural properties of node $v_i$.

The recent advances on GE-based GLRSs have been largely influenced by the skip-gram model (Mikolov et al. 2013a) designed originally to learn word representations w.r.t. the words context in a sequence e.g. a sentence. For a specific type of graph, skip-gram can be used on path sequences extracted from the graph in which nodes can be analogous to words, and paths can be analogous to sentences. Inspired by this, a series of graph distributed representation-based GLRSs using skip-gram-related algorithms, e.g. DeepWalk (Perozzi et al. 2014), LINE (Tang et al. 2015), and Node2vec (Grover and Leskovec 2016), gradually emerged and achieved encouraging success (Verma et al. 2019; Palumbo et al. 2017; Jiang et al. 2018b; Vijaikumar et al. 2019; Gao et al. 2018; Ali et al. 2020; He et al. 2016; Wang et al. 2020a; Gharahighehi et al. 2021; Fu et al. 2020). For instance, the authors of Gao et al. (2018) apply DeepWalk which aims to maximize the average logarithmic probability of all vertex context pairs in a random walk sequence, to learn user and item representations on a multi-source homogeneous item graph to consider item structure, textual content and tag information simultaneously which are then used for collaborative filtering. In Palumbo et al. (2017) the authors generate user and item representations with Node2vec, an extension of DeepWalk by leveraging a biased random walk to navigate the neighbourhood nodes, on a complex heterogeneous knowledge graph, which are then used to compute property-specific relatedness scores between users and items as the input for the learning to rank approach, resulting in optimizing top-N item recommendations.

Another research line of GE-based technique adopts translation-based embedding models inspired by Mikolov et al. (2013b), e.g. TransE (Bordes et al. 2013). Different from DeepWalk-related approaches, TransE explicitly models entities and relationships among entities into the same space or different spaces while preserving certain information of the graph, which is later generalized into a hyperplane translation (TransH (Wang et al. 2014)) and a translation in separate entity space and relation spaces (TransR (Lin et al. 2015)). The basic idea behind TransE is that the relationship between two entities corresponds to a translation between the embeddings of entities, that is, $\boldsymbol{h} + \boldsymbol{r} = \boldsymbol{t}$ where $\boldsymbol{h}$, $\boldsymbol{t}$, and $\boldsymbol{r}$ represent head entity, tail entity and relation between $\boldsymbol{h}$ and $\boldsymbol{t}$ in triplet $(h, r, t)$ in a graph. Researchers attempted to adopt such translation-based models for e.g. a knowledge graph embedding for recommendation (Ai et al. 2018; Wang et al. 2020b; Chen et al. 2019d; Cao et al. 2019; Wang et al. 2018b, 2019a). For example, Wang et al. (2020b) assign a basic representation and various relational ones for each item from a directed homogeneous graph via TransE, which are then combined dynamically by temporal kernel functions, providing both recommendations and explanations. Chen et al. (2019d) adopt TransH to embed the objects' social relationships from the homogeneous graph into a shared lower-dimensional space and learn a user's dynamic preference via a probabilistic model from the user–item bipartite graph. Finally, the recommendation list is generated with item-based collaborative filtering.

### 5.4 Deep learning-based techniques

Deep learning (DL) has driven a remarkable revolution in recommender applications as can be seen by the number of research publications on deep learning-based recommendation techniques having increased exponentially recently. It has been applied to multiple types of graphs from single graph to multiple graphs, from homogeneous to heterogeneous graphs from different recommendation domains. To draw an overall concept of this field, we further classify the existing DL-based approaches into DNN, auto-encoder, attention mechanism, reinforcement learning, graph neural network, transformer-based approaches and deep hybrid models as shown in Fig. 7.

#### 5.4.1 Deep neural network (DNN)

A deep neural network (DNN) is adopted to model complex non-linear relationships with generated compositional models where the object is expressed as a layered composition of primitives. By piling up layers, composition of features from lower layers can be extracted and learned (Bengio 2009). DNNs are typically feedforward networks in which data flow from input layer, are transformed into vector representations, and projected into a different space to an output layer without looping back. DNN-based techniques can be used for tree graphs (Huang et al. 2019) and complex heterogeneous graphs (Sun et al. 2018; Mezni et al. 2021; Wang et al. 2019g) in e-commerce and entertainment domains, or homogeneous graphs (Wu et al. 2019a) in social network domain. Taking graphs as input, DNN can learn high-order interactions among nodes by stacking layers with non-linear transformations (Wang et al. 2017; Wu et al. 2019a). For instance, Wang et. al (Wang et al. 2017) apply several multiple layers with a pooling operation to explore interaction patterns between user, item and their attributes from multiple cross-domain graphs taking nodes' one hot encoding as input. To learn the propagation effect of social influence of users, they further employ the smoothness regularization term to closely embed users connected by social networks from different data sources.

Another variation in DNN is Recurrent neural networks (RNNs) (Cho et al. 2014; Hochreiter and Schmidhuber 1997). The original superiority of RNNs can well capture the dependencies among items from graphs in time-sensitive user–item interaction sequences or in session-based recommendation settings. However, the limitations lie in that it is difficult to model dependencies in a longer sequence, and training is burdened with high cost, especially with the increase in sequence length. Thus, some works combine RNN with other mechanisms to balance this disadvantage of RNN. For instance, Huang et al. (2019) design a memory-module to extract a user's fine-grained preference on a taxonomy from multiple hops reasoning on a tree-based graph, together with a GRU layer to learn the sequential pattern. Wang et al. (2019g) make it different by adopting an LSTM layer to model the sequential dependencies of entities and relations on a complex heterogeneous KG, generating path representations followed by a pooling operation to obtain a prediction signal for user–item pairs. Besides, using RNN for long sequence modelling also suffers from the vanishing and exploding gradient problem because of the choice of the number of layers and the activation functions, which is a common problem in many types of neural networks, e.g. feed-forward neural network, and CNN. Despite its limitations, the RNN-based approach still dominates in sequential recommendations due to its recurrent nature that matches the natural way of our brain to read one after another in a sequence mode.

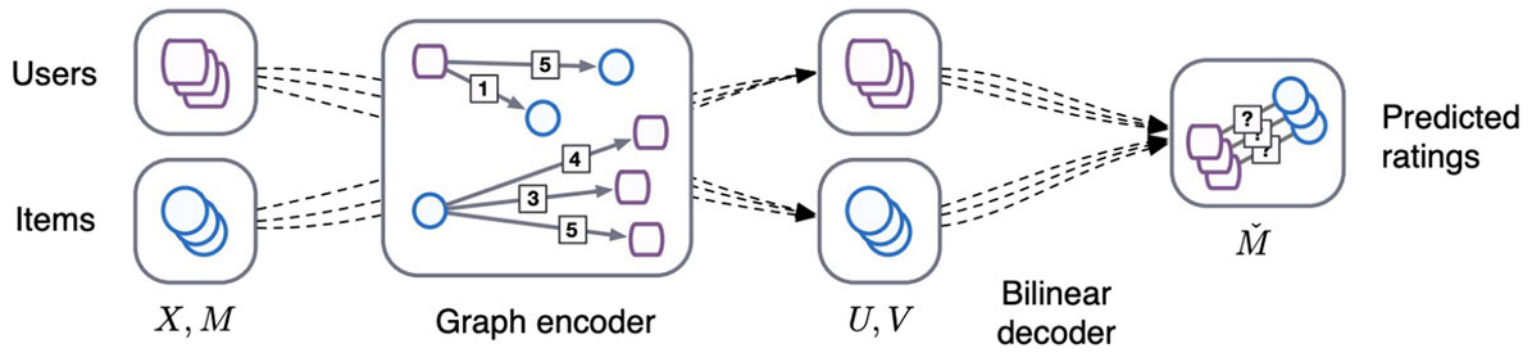

**Fig. 10** Illustration of the graph auto-encoder framework in GLRS (Berg et al. 2017).The system input bipartite user–item graph which is represented an input data source $X$ and adjacency matrix $M$. Then, the graph encoder learns node representations of users and items $U$, $V$, which go through the decoder to derive the predicted rating matrix $\hat{M}$

Convolutional neural networks (CNNs) (Krizhevsky et al. 2012) are capable of extracting local and global representations from heterogeneous data sources such as textual and visual information. To leverage CNN for extracting graph structured data, Wang et al. (2018c) extend traditional CNN which allows flexibility in incorporating symbolic knowledge from a complex heterogeneous knowledge graph for learning sentence representations.

### 5.4.2 Auto-encoder

A basic auto-encoder (AE) contains an encoder which encodes (projects) high-dimensional inputs $\boldsymbol{X}$ to low dimensional hidden representations $\boldsymbol{Z}$, and a decoder which decodes (re-projects) hidden representations $\boldsymbol{Z}$ to the output $\hat{\boldsymbol{X}}$ that looks like the original input $\boldsymbol{X}$. The objective is to minimize the reconstruction error, and find the most efficient and informative compact representations for the inputs. In most GLRSs studies, AE is applied to learn a complex heterogeneous graph (Zhang et al. 2016) in the entertainment and book recommendation domains.

To apply AE to graph-structured data for a recommendation purpose, Zhang et al. (2016) first use TransE to learn graph topological information from the complex heterogeneous knowledge graph. Then, stacked denoising auto-encoders and stacked convolutional auto-encoders are adopted to learn textual and visual representations of items, which are the input for the collaborative filtering framework. Later, Berg et al. (2017) consider a recommender system as a matrix completion task, and propose to apply a graph auto-encoder to produce latent features of user and item nodes through a form of message passing on the bipartite user–item interaction graph. The learned latent user and item representations are used to reconstruct the rating links through a bilinear decoder. Generally speaking, a graph auto-encoder takes node feature embedding $\boldsymbol{X}$ and adjacency matrix $\boldsymbol{A}$ as inputs, generating latent variable $\boldsymbol{Z}$ as output through the encoder (inference model). To reconstruct the graph structure data, the decoder (generative model) takes $\boldsymbol{Z}$ as input and outputs a reconstructed adjacency matrix $\hat{\boldsymbol{A}}$. Based on Berg et al. (2017), Zhang et al. (2019a) go a step further by proposing a new stacked and reconstructed graph convolutional network for a user–item bipartite graph, which takes low-dimensional user and item embeddings as the input to the model and solves the cold start problem by reconstructing the masked node embeddings with a block of graph encoder-decoder in the training phase. Figure 10 illustrates how an auto-encoder operates on graph-structured data for a recommendation purpose. The problem of the auto-encoder framework is that it usually leads to a local optimum due to the back-propagation algorithm it employs (Tian et al. 2014), which is also the common problem of most deep learning-based techniques that adopt back-propagation as a training procedure. Besides, the encoder-decoder architecture requires that the complete sequence of information must be captured by a single vector, which poses problems in holding on to information at the beginning of the sequence and encoding long-range dependencies.

### 5.4.3 Attention mechanism

The attention mechanism (Bahdanau et al. 2014) is motivated by human visual attention. For example, people only need to focus on specific parts of visual inputs to understand or recognize them. The attention mechanism is proposed to determine the significance of the inputs e.g. sequences. The effectiveness of the attention-based techniques in RSs has been verified and aroused considerable attention over recent years. In attention-based GLRSs, inputs are weighted with attention scores and outputs are normally vectors that combine different importance of the inputs. Attention mechanism can be used to allow the learning process to focus on parts of a graph that are more relevant to a specific task. Generally, it can be used in conjunction with MLP, CNN, RNN and other deep learning-based architectures. Thus, the heart of the attention-based techniques is how to obtain and calculate the attention weight of each input part. In this paper, attention mechanism can be used for multiple different types of graphs in various recommendation domains. For instance, in the e-commerce domain, the Amazon dataset can be formed into a tree graph (Gao et al. 2019b), k-partite graph (Wang et al. 2019d) and complex graph (Han et al. 2018) learned with an attention network. The Ciao, Epinions, Taobao and Kuaishou datasets can be formed into homogeneous graphs (Chen et al. 2019a; Chang et al. 2021b) learned with attention mechanisms. Data sources in the entertainment domain can be constructed into k-partite graphs (Wang et al. 2019d; Xin et al. 2019), complex heterogeneous graphs (Han et al. 2018; Wang et al. 2019a), homogeneous graphs (Hao et al. 2021b; Chen et al. 2019a) and multiple graphs (Xia et al. 2021a) learned with an attention network. Attention mechanisms can also be applied in POI (Wang et al. 2019d; Hao et al. 2021b) and book (Wang et al. 2019a) domains.

There are three attention mechanisms commonly used in recent studies: (1) the *vanilla attention* mechanism learns the attention scores for the input data by transforming the representations of input data via fully connected layers, and then adopting a softmax layer to normalize the scores (Han et al. 2018; Wang et al. 2019d; Gao et al. 2019b). Han et al. (2018) propose to use multi-layer MLP to learn a user/item aspect-level representation based on extracted meta-paths from the complex heterogeneous graph. Then an attention mechanism is adopted to weigh the contribution of different aspect-level latent factors to final user/item representations. (2) The *self-attention* mechanism (Vaswani et al. 2017) gained exposure recently as it can replace RNN and CNN in sequence learning, achieving better accuracy with lower computational complexity. It focuses on the self-matching of a sequence whereby the attention weights are calculated by the multiplication between key and query vectors transformed from the input sequence (Cen et al. 2019; Wu et al. 2019b). For instance, Cen et al. 2019 adopt self-attention to capture the influential factors between different edge types of the neighbours of a specific node on the attributed bipartite graph. (3) The *co-attention* mechanism focuses on co-learning and co-matching of two sequences whereby the attention weights of one sequence are conditioned on the other sequence, and vice versa. Some studies prefer to classify co-attention and self-attention as one category (Zhang et al. 2019b; Sun et al. 2019), but for clarity, in this survey we describe them separately. In Xu et al. (2019b), the authors design a parallel co-attention mechanism to dynamically infer the primary reasons of the user purchase decision, assigning higher attention weights to more relevant meta-paths extracted on the k-partite graph. Other studies adopt attention variations based on these three categories. For instance, in Verma et al. (2019) the authors adapt a skip-gram to a merged heterogeneous user–item interaction and use social networks followed by a multi-layer and multi-head attention (Vaswani et al. 2017) mechanism to learn the different importance of entities. Multi-head attention improves self-attention mechanisms to draw global dependencies between inputs and outputs by eschewing the use of recurrence in neural network and running through an attention mechanism several times in parallel. In Liu et al. (2020) the authors adopt Sentence-BERT (Reimers and Gurevych 2019), a language model that is based on multi-head attention and a bidirectional training procedure, to explore the potential links between item based on reviews. The learned item representations from BERT are then used to generate an item subgraph according to the cosine similarities between all items.

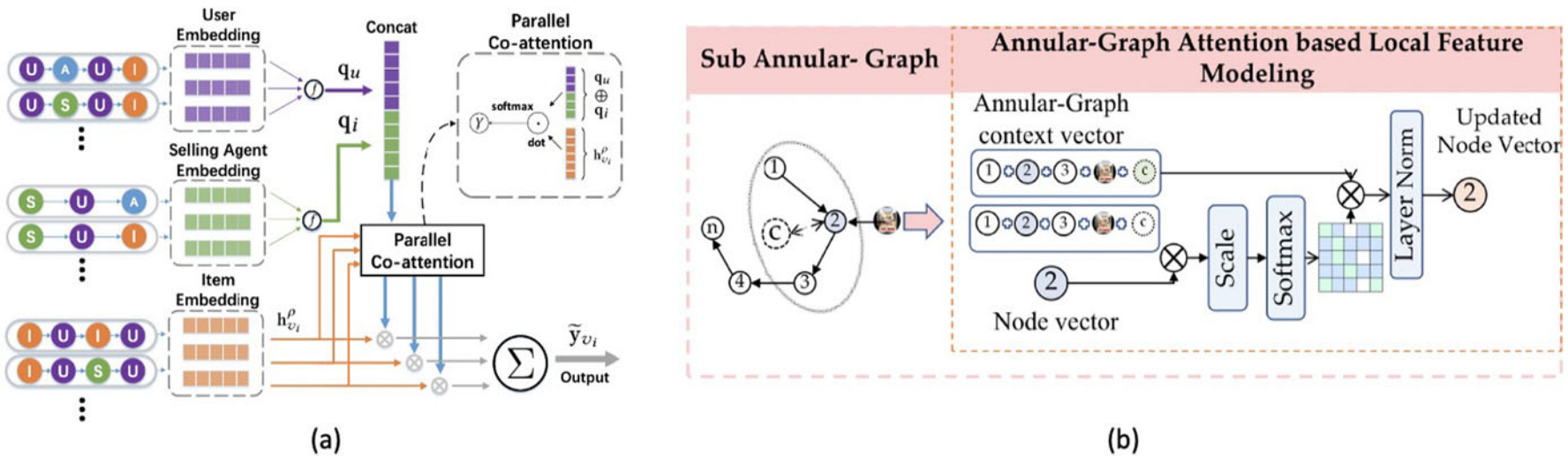

**Fig. 11** Illustration of different attention mechanisms in GLRS. **a** In Xu et al. (2019b), the co-attention component takes embeddings derived from different meta-paths as inputs. Then the query vectors $q_u$, $q_i$ transformed from user and selling agent embeddings, as well as the item embeddings $h^{\rho}_{v_i}$ go through the parallel co-attention network to learn the item embedding $\tilde{y}_{v_i}$. **b** To update the representation of the ring node 2 in the context of the annular-graph, a self-attention mechanism takes two context vector and the node 2 vectors as input to learn the importance of context nodes to the central node (Hao et al. 2021b)

Figure 11 illustrates several above mentioned attention mechanism architectures for GLRSs. The core of the attention mechanism of focusing on the most relevant parts of the input by providing a direct path to the input helps to alleviate the bottleneck problem of the vanishing gradient and to resolve the disadvantage of the encoder-decoder architecture that has the problem of remembering long sequence dependencies. However, one believes that the attention mechanism adds more weight parameters to the model, which increases training time, especially for long input sequences.

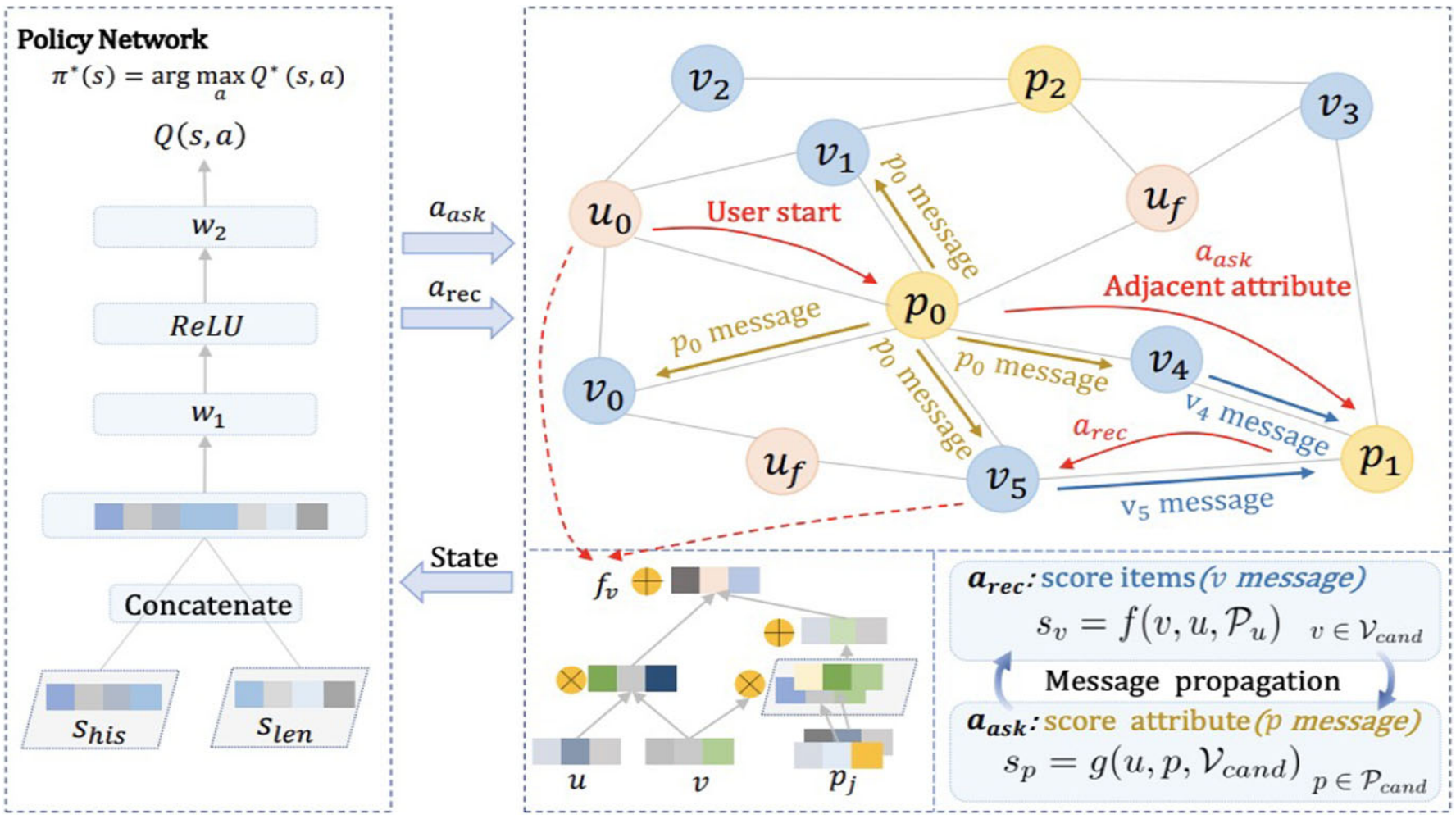

**Fig. 12** Deep reinforcement learning-based GLRS with a knowledge graph (Lei et al. 2020a). The system first performs the walking starting with the target user $u_0$ over the adjacent attribute vertices on the complex heterogeneous graph, resulting in a path to the desired item. Two reasoning function $f$ and $g$ score attributes and items. Then, the policy network takes the state vector $\boldsymbol{s}$ as input and outputs the values $Q(\boldsymbol{s}, a)$, indicating the estimated rewards for two actions $a_{ask}$ and $a_{rec}$

### 5.4.4 Deep reinforcement learning (DRL)

Reinforcement learning (RL) uses a trial-and-error experience with an agent that learns a good behaviour by modifying or acquiring new behaviours and skills incrementally. During such a learning process, the agent interacts with the environment and must make value judgements so as to select good actions over bad. Actions that get them to the target outcome are rewarded (reinforced). Deep reinforcement learning (Mnih et al. 2015) goes a step further by incorporating deep neural networks to represent the knowledge acquisition progress. It has been mainly adopted to learn k-partite (Song et al. 2019a) and complex heterogeneous graphs (Lei et al. 2020a; Liu et al. 2021a; Xian et al. 2019) in entertainment, book and e-commerce domain for recommendation purposes.

In GLRSs, one can take a path generation procedure as a decision-making process for training with RL, so that the optimal recommendation results as well as the interpretation of the results can be generated at the same time (Song et al. 2019a; Xian et al. 2019; Lei et al. 2020a; Wang et al. 2020g). For instance, Xian et al. (2019) propose to use an RL approach where an agent starts from a given user, and learns to navigate to the potential items of interest on the complex heterogeneous KG. After that, the rea-soning path history can serve as a genuine explanation for the recommendation results. Similarly, Song et al. (2019a) formulate the generation of user-to-item paths extracted from a k-partite graph as a sequential decision process. Specifically, it defines the target user as the initial state and then walks on the constructed heterogeneous user–item–entity graph as actions. In the work of Zhou et al. (2020b), the authors adopt a complex heterogeneous KG to improve the sample efficiency as well as interactive recommendation performance by applying a deep Q-network to fit on samples from the local graph of the KG rather than the whole graph. Interestingly, we can find that most DRL-based recommendation approaches utilize a KG as an important medium to learn user-to-item inference. Figure 12 illustrates a typical example of adopting a KG with DRL for recommendation purpose. It is probably due to the explicit association between the target user and items which reveals the user's potential interests, compared with traditional recommendation systems, a KG-based recommender system can mine more potential relationships between nodes for learning user and item representations.

DRL-based approaches have great potential in decision-making and long-term planning in a dynamic environment (Silver et al. 2016). However, the ideal way to train a DRL model to learn the optimal recommendation policy is to train the agent online, which cannot always be satisfied. One commonly used training strategy is to make use of offline logged data directly, but it will suffer from the estimation bias problem under the real-time interaction setting (Chen et al. 2019c). Besides, similar to other deep learning-based techniques, DRL-based approaches also lack interpretability. More importantly, few appropriate platforms or resources for developing and testing DRL-based techniques in academia exist (Fang et al. 2020).

### 5.4.5 Graph neural networks

Graph neural network (GNN) enjoys a massive hype as recent works have witnessed a boost of performance in RSs. They are motivated from CNN and graph embeddings and designed specifically on graph-structured data in the non-Euclidean domain (Zhang et al. 2020). GNN can be applied from homogeneous graphs (Zhu et al. 2021b; Isufi et al. 2021), k-partite graphs (Wang et al. 2019f; Chen et al. 2020a; Wang et al. 2019e; Sun et al. 2020; Wu et al. 2021b; Fan et al. 2019a; Chen et al. 2021a; Ying et al. 2018; Wei et al. 2021; Li et al. 2021a) to complex heterogeneous (Zhao et al. 2019; Zheng et al. 2021; Wang et al. 2019c, b; Wu et al. 2021a; Zhang et al. 2021d) and multiple ones (Zhu et al. 2021a; Zhang et al. 2021c; Liu et al. 2020; Huang et al. 2021b; Tang et al. 2021; Wu et al. 2019b; Tian et al. 2021; Chang et al. 2021a) in different recommendation domains.

GNN achieve improvements in recommendation results by capturing the higher-order interaction in user–item relationships through iterative propagation resulting in better user/item representations. Specifically, GNN aim to iteratively aggregate feature information from neighbours and integrate the aggregated information with the current node representation (Wu et al. 2020b). Further, they can simultaneously model the diffusion process on the graph with the RNN kernel. Following the existing work of Wu et al. (2020a), we categorize GNN as spectral (Zheng et al. 2018; Wang et al. 2020a; Farseev et al. 2017) and non-spectral approaches (Sun et al. 2020; Isufi

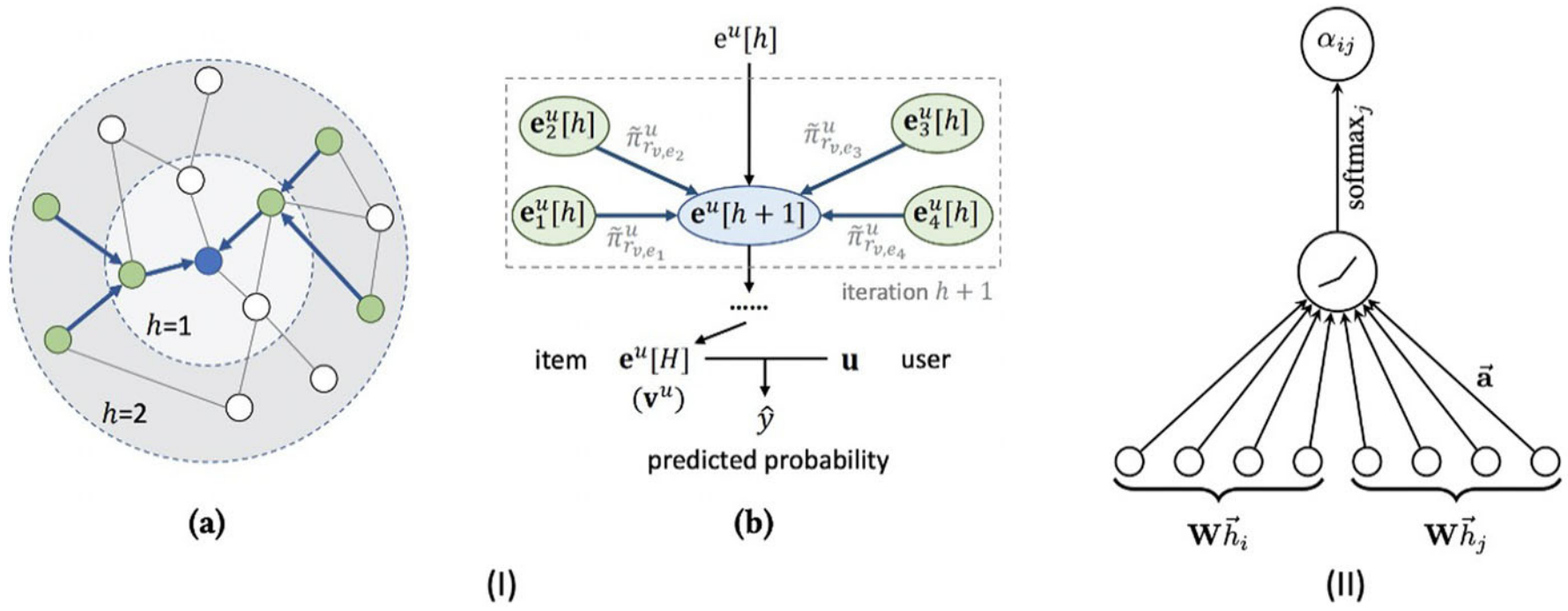

**Fig. 13** Illustration on GNN-based GLRSs. (I) The representation of the central node updates by incorporating the influence of its neighbourhood representation in GNN algorithm (Wang et al. 2019c). (II) GGAT takes into account the different effects of neighbour nodes on the central node, and combines the attention mechanism with the GNN node propagation process to update the representation of the central node (Veličković et al. 2017)

et al. 2021; Kim et al. 2019; Chen et al. 2020a; Wang et al. 2019c, f, b; Fan et al. 2019a; Liu et al. 2020; Ying et al. 2018; Zhao et al. 2019). Spectral GNN are based on spectral graph theory (Shuman et al. 2013) which studies connections between combinatorial properties of a graph and the eigenvalues of matrices associated with the graph, e.g. laplacian matrix. They focus on the connectivity of the graph rather than geometrical proximity. For instance, Farseev et al. (2017) performs spectral clustering to form user community w.r.t. user side information e.g. geographical regions, user's active timestamp from a complex heterogeneous graph, which are then considered to sort all candidate items to a generate ranked list for recommendation. Zheng et al. (2018) propose to use a spectral convolution operation in the spectral domain of the bipartite user–item graph to alleviate the cold-start problem of RS.

The non-spectral approaches mainly include aggregator and updater to learn a multi-layer graph. The aggregator is responsible for collecting information from neighbourhood nodes and related edges, while the updater aims to merge the propagation information around the central node and collected through the aggregator. Normally, GNN are utilized to learn the representations of nodes and links of the graphs, which are then used for the following recommendation strategies, e.g. rating prediction and link prediction etc. For instance, Monti et al. (2017) propose a GCN-based technique for recommender systems for the first time, in which GCN is a variant of GNN and used to aggregate information from two auxiliary user–user and item–item homogeneous graphs with the convolutional operation. The latent factors of users and items were updated after each aggregation step, and a combined objective function of GCN and MF was used to train the model. Ying et al. (2018) propose to use GCN to generate item embeddings from both a bi-partite graph structure as well as item feature information with random walks for recommendations. It can be applied to very large-scale web recommenders and has been deployed in Pinterest to address a variety of real-world recommendation tasks. In Chen et al. (2019a) the authors adopt a GNN layer for modelling both the local and global influence of user social relations on constructed homogeneous user social graph. Graph attention networks (GATs) (Veličković et al.

2017) are an enhanced version of GNN which utilize masked self-attention layers to limit the shortcomings of prior graph convolutional-based approaches. An attention weight $\alpha_i \in [0, 1]$ is assigned to the neighbourhood nodes of a target node $n_t$, where $\sum_{i \in N(t)} \alpha_i = 1$ and $N(t)$ denotes the set of neighbouring nodes of $n_t$. One advantage of applying attention to graphs is to avoid the noisy part of a graph so as to increase the signal-to-noise ratio in information processing. Specifically, GAT aims to compute the attention coefficients

$$\alpha_{ij} = \frac{exp(LeakyReLU(\overrightarrow{a}^T[W\overrightarrow{h_i}||W\overrightarrow{h_j}]))}{\sum_{k \in N_i} exp(LeakyReLU(\overrightarrow{a}^T[W\overrightarrow{h_i}||W\overrightarrow{h_k}]))} \tag{3}$$

where $\overrightarrow{a}$ and $W$ is the weight matrix. $h_k$ is the neighbour node embedding of node $n_i$ whose node embedding is $h_i$. Figure 13(II) illustrates the schematic diagram of the attention operation of GAT.

Despite their verified effectiveness in the community of graph-based recommendations, they suffer from the expensive computation overhead with the exponential growth of the neighbourhood size as the layers stacked up (Ying et al. 2018). Besides, researchers empirically show that the performance of GNN quickly degenerate when the number of layers is deep owing to that the effectiveness of informative neighbours will be diminished in large amount irrelevant neighbours (Liu et al. 2019b). To solve this, Xu et al. (2019b) design a relation-aware GNN with an attention mechanism to prioritize neighbours based on their importance. Then a meta-path defined receptive field sampler is integrated to derive the node embeddings as well as address the rapid growth of the multiple-hop neighbourhood of each node from a k-partite graph, which is followed by a co-attention mechanism for differentiating purchase motivations. In Zhang et al. (2019a), the authors also point out that training GCN-based models for rating prediction faces the label leakage issue, which results in the overfitting problem and significantly degrades the final performance, which can be improved by removing the sampled edges. Figure 13a, b illustrates examples of how GNN can be used for recommender systems in Wang et al. (2019c), and the main component of GAT in Veličković et al. (2017), respectively.

Although some problems have been proposed by researchers for improvement or solutions, other challenges still exist and deserve more attention from both academia and industry. First, though GNN-based graph recommendation strategies can incorporate a high-order proximity of vertices, it suffers from the problem of performance degradation and complexity increase with the increase of the number of layers. As a result, it is more prone to encounter the over-smoothing problem with the increase in the network depth (Liu et al. 2019b; Li et al. 2018a; Yu et al. 2021). This phenomena can become a pervasive problem when learning large-scale graph/networks when aggregating high-order information from distant neighbours is necessary. Second, current GNN are mainly applied for a static graph, but how to apply GNN for dynamic graphs with changing structures is still an open challenge.

### 5.4.6 Deep hybrid models

In order to deal with more complicated and diverse problems, as well as process more complex graphs, many graph-based recommendation models utilize more than one deep learning technique. The flexibility of neural blocks in deep neural networks makes it possible to combine several neural components to complement one another and form a more powerful hybrid model. The use of a variety of different DL-based components can also maximize the strengths and improve the defects of a single technology to a certain extent. Such hybrid models can be leveraged to learn homogeneous graphs (Song et al. 2019b; Abugabah et al. 2020; Xu et al. 2019a; Wu et al. 2019c; Huang et al. 2021a), k-partite graphs (Kim et al. 2019; Zhang et al. 2019a; Wang et al. 2020f; Liu et al. 2021c; Xie et al. 2021; Xu et al. 2019b; Xia et al. 2021b), complex heterogeneous graphs (Wang et al. 2020g; Xie et al. 2021; Yang and Dong 2020; Zhou et al. 2020b; Sang et al. 2021; Sheu and Li 2020; Shi et al. 2021; Wang et al. 2018c; Zhou et al. 2020a; Yang et al. 2021a), hypergraphs (Wang et al. 2020c; Gharahighehi et al. 2020), and multiple graphs (Monti et al. 2017; Zhang et al. 2021a; Xia et al. 2021c; Fan et al. 2019b; Liu et al. 2021b; Monti et al. 2017) in social network, e-commerce, entertainment, academic, book and many other recommendation domains.

In Sun et al. (2018), the authors employ a batch of bi-directional recurrent networks (Schuster and Paliwal 1997) to learn the semantic representations of each path extracted from complex heterogeneous KG. Then an attention gated hidden layer is applied to learn the different importance of the derived paths between two entities followed by a pooling operation and a fully connected layer for rating prediction. Zhang et al. (2019a) propose to leverage multi-link GCN as an encoder and two-layer feedforward neural network as a decoder to learn the user and item (users and items are denoted as nodes) representations, considering both node's content information as well as structural information of the undirected user–item bipartite graph. Some studies leverage different techniques to learn various graph features, such as node attributes and graph structure. Zhang et al. (2016) construct a complex heterogeneous knowledge graph to learn a user's potential preferences, where the item nodes associate with textual and visual features as their attributes. To model such multi-modal information, the authors first apply a network embedding (TransR) approach to extract items' structural representations by considering the heterogeneity of both nodes and relationships, followed by a stacked denoising auto-encoder and stacked convolutional auto-encoder to extract items' textual and visual representations, respectively. Finally, the pair-wise ranking between items is considered to learn the CF architecture. Deep hybrid approaches have become a trend in solving complex recommendation problems, facing the complicated and changeable network structure for modelling dynamic user preferences.

## 5.5 Discussion of graph-based recommendation models

In this section, we present the main ideas and the basic technical details of each class of graph learning-based recommendation approaches. From the descriptions above, we make several observations: (1) traditional graph learning-based approaches may suffer from information loss, e.g. nearest neighbour-based approaches. Some of them ignore long-term or high-order dependencies such as the latent factor model (Tao et al. 2021). However, they laid a theoretical foundation for the later development of graph learning-based technologies, so that many recent advanced deep learning-based techniques for graphs still use traditional algorithms as the basic framework. (2) For path-based approaches, they either rely on domain knowledge which may not always be applicable e.g. meta-path-based similarities, and/or require explicit path reachability which may incorporate noisy, meaningless paths and thus do not always have a positive impact on recommendation results (Noia et al. 2016). Besides, recommendations that rely on similarity measures cannot be easily applied to large-scale networks. (3) Graph embedding-based approaches pave the way for more complex and high-order features among nodes and links modelling. More and more state-of-the-art GLRSs leverage GE combined with deep learning approaches such as an attention network for more efficient recommendation tasks. (4) With the many achievements of the deep learning-based GLRSs, the number of research works in the field has become exponentially increased. Deep learning-based approaches can be applied to more complicated graphs with multi-type nodes and links, as well as additional attributes associated with graphs (Song et al. 2019b). Besides, deep learning-based approaches are more robust to sparse data and can adapt to the varied magnitude of the input (with the help of e.g. attention mechanism) (Fang et al. 2019). However, interpretability and efficiency are still the main concerns for most GE-based and deep learning-based GLRSs which need to be further studied in the future. (5) From Table 5, by relating graph types to their associated modelling technologies, we observe that tree graphs are modelled mainly by traditional techniques, while other types of graphs are modelled and learned mainly through deep learning and graph embedding-based approaches. Besides, attention mechanisms become especially prevalent and are adopted for nearly all types of graphs for selecting branches, filtering noisy nodes, and learning better nodes and edges representations for recommendation purposes.

## 6 Challenges and open issues in GLRS

Graph-based recommendation is an exciting and rapidly growing research area that attracts attention from both industrial and academic domains. While existing works have established a solid foundation for GLRSs research, this section reveals several challenges and promising prospective research directions. Specifically, there are two types of challenges: (1) Challenges still unsolved by graphs, which include explainability, fairness and generality issues; (2) Challenges caused by graphs and their limitations, including scalability, dynamic graph, and complex heterogeneity learning issues. We will explain these issues separately.

**Explainability on graphs.** A good explanation for recommendation results can help to improve the transparency, persuasiveness, effectiveness, trustworthiness and satisfaction of recommender systems, facilitate system designers for better system debugging, and allow a user to control how the system utilizes her profiles making the RS scrutable. Earlier studies provide explanations of GLRSs highlighting top keywords or aspects as an explanation for recommendation results (He et al. 2015), but none of them shows the constructed graphs.

With the surge of deep learning-based approaches in GLRSs, it is even harder to provide convincing explanations and calibrate why the recommendation models are effective and thus yield a robust model for varied scenarios. Ma et al. (2019) provide recommendation explanations according to the learned reasoning rules on heterogeneous graphs with ground-truth item associations in the knowledge graph. The emergence of attention mechanisms has more or less eased the non-interpretable concerns of deep learning-based recommendations on graphs. The learned attention weights can tell which parts of the input graph contribute more than others with higher attention scores to the recommendation results. However, the faithfulness of the higher attention weights in contributing to the performance of the recommendation is still doubtful (Liu et al. 2022).

In addition, most existing graph leaning-based recommendation models generally consider paths along with nodes in the KG as pertinent signals for recommendation (Hu et al. 2018). User–item paths in the KG can directly serve as explanations that provide the reason why an item is recommended (Xian et al. 2019). In (Fu et al. 2020), Fu et al. considered a relational structure serving as explanation across different paths with similar semantics w.r.t. relations in each path. We argue that this lacks reliable evaluation metrics for the explainability and persuasiveness for target users, such as GLUE (Wang et al. 2018a) for language understanding. For some kinds of graphs, e.g. hypergraphs and multiple graphs, it is more difficult to provide explanations with such graphs due to the dispersion of edges or nodes. Besides, how to show user dynamic preferences on graphs is also worth studying. Besides, we need to consider the target population who we provide explanations for. This can be the end-users or researchers. For both groups, whether they are satisfied with this explanation and whether this explanation is sufficient still requires further empirical verification. Such verification cannot just stay on a small-scale user study or case study. It requires more target audience participation and different situations should also be taken into account.

**Fairness.** Fairness in recommendation has gradually attracted increasing attention in recent years and has been studied mainly as an equity and parity problem for individuals or groups of users. Current RSs discriminate unfairly among the users in terms of recommendation performance, and further, the systems may discriminate between users in terms of explanation diversity (Farnadi et al. 2018). One reason is probably related to the issue of data imbalance. E.g. economically disadvantaged customers tend to make fewer purchases, leading to imbalanced data. However, such imbalances may lead to biased models that exhibit unfairness w.r.t. the recommendation quality and explanation (Fu et al. 2020). Though active users tend to interact with more items, an empirical study (Fu et al. 2020) shows that the majority of users are inactive users who are easily disregarded by recommendation engines. The imbalanced data can easily lead to biased observation on graphs where path inference from user-to-item usually participates in the process of graph learning and meanwhile provide recommendation explanations. Many researches have been done to alleviate data sparsity issue. Farnadi et al. (2018) propose to solve the fairness on the user side for both the individual- and group-level for KG enhanced explainable RSs. Specifically, they reveal that the unfairness issue is due to data imbalance through an empirical study on e-commerce dataset, namely Amazon. Then they propose fairness metrics in terms of path diversity as well as recommendation performance disparity based on KG. In Gharahighehi et al. (2021), the authors constructed a hypergraph taking into account multiple stakeholders in the news domain to mitigate the imbalance problem caused by stakeholders with few articles. Exposure bias can be caused by the users being only aware of a very small fraction of items in a large dataset. Chen et al. (2019a) pointed out the exposure bias that users are only aware of a very small fraction of items in a large dataset so that they infer data confidence with the help of users' social network and draw different weights on training instances via personalized random walk to alleviate it. Another possible reason lies in the incomplete evaluation metrics, which renders the inclination of the learning objectives on accuracy driven for recommendation purposes.

Despite some studies considering the influence of fairness in GLRSs, related research is still quite limited with many issues remaining to be focused on. For instance, whether the construction and learning process of the graphs affects the fairness and discrimination of the ranking of recommendation results, whether there exists algorithmic bias among various graph learning technologies, and whether fairness, bias, and discrimination conflict with the accuracy of graph-based recommendations. Some biases can be dynamic. For instance, in the real world, users' preferences, exposure, and relations may evolve over time (Chen et al. 2019a). How to solve the dynamic bias problem in GLRS is also one of the research directions worth thinking about in the future.

**Scalability on large-scale graphs.** Scalability is an essential factor that affects the applicability of recommendation models in real-world scenarios. To deal with large-scale graphs, most existing models choose to adopt a sampling scheme to construct subgraphs following the sampling strategy proposed in GraphSage (Hamilton et al. 2017). Some use the random walk strategy to get the neighbourhood nodes and links, while others consider using the shortest path algorithm for subgraph construction. Another algorithm to increase model scalability is to use a clustering scheme. Whether using sampling or clustering, a model will lose part of the graph information. The scalability is gained at the price of corrupting graph completeness. A node may miss its influential neighbours with a bad sampling strategy, and a graph may be deprived of a distinct structural pattern by clustering. Though the subgraph strategy makes GNN-based algorithms applicable no matter how large-scale the whole graph is, the shortcoming is that the node representation should be recalculated for each propagation layer. Thus, how to tradeoff the algorithm scalability and graph integrity could be one of the further research directions. More researches can be studied on the sampling strategy in integrating more informative information from neighbourhood nodes and links while minimizing the harm to the graph integrity. Recently, Kyriakidi et al. (2020) propose to adopt graph databases as a base to improve the data scalability and meanwhile build recommendation models on top. Different from other graph-based recommender systems which focus on model complexity when considering the efficiency problem, the work of Kyriakidi et al. (2020) transforms the recommendation problem into a path optimization problem from start nodes to end nodes on heterogeneous graphs, which sheds light on a new perspective for improving the scalability of GL-based recommendations.

**Data Sparsity on Graphs.** Data sparsity issues will cause the graph's adjacency matrix to be sparse and affect recommendation performance. In order to alleviate this problem, one way is to construct auxiliary graphs by mining contextual information of users or items, such as friendship connections amongst users, co-purchase networks associated with products and services to the end-users, or trust relationships associated with users, entities associated with items. On the other hand, data augmentation techniques can be used. The former has been widely used and studied (Wu et al. 2019b), while the latter has not been extensively explored. Recently, some studies try to alleviate the data sparsity problem with item/segment dropout to augment data (Wu et al. 2021b), with which edges are randomly dropped out in the constructed graph, resulting in the robustness of recommendation models facing noisy input data such as unsatisfied clicking or viewing behaviour. However, it may also lead to sparser data. In Xia et al. (2021c), the authors adopted a self-supervised graph co-training strategy for learning session representations with two different encoders in the session-based recommendation. As the initial attempt in many graph learning-based recommendation domains, more in-depth attempts are still needed, and the advantages and disadvantages of different data augmentation technologies with different types of graphs in different recommendation domains also require a comprehensive analysis.
**Recommendation on Dynamic Graphs.** Most research focuses on the modelling and learning of static graphs, but neglects the dynamic properties of graphs. In many real-world scenarios, the graphs changes over time, such as people may be followed or unfollowed day to day in a social network, or newly published news articles appear everyday resulting in changes of news-related graphs for news recommendation. As a result, some works crop the dynamic graph as a sequence of graph snapshots (Abugabah et al. 2020; Xia et al. 2021d; Huang et al. 2021a), most of which are in session settings. Though a surge of works consider modelling dynamic graphs from the changes of the adjacency matrix (Li et al. 2017), leveraging dynamic random walk sampling (Nguyen et al. 2018), Hawkess process (Huang et al. 2020) or combining with other algorithms capturing the dynamic properties of the graph (Xu et al. 2020), they do not adapt to recommendation scenarios. Thus, we believe it is an important research direction with significance and practical value.
**Complex Heterogeneity Learning.** Apart from the user–item bipartite graph, most heterogeneity of graphs is reflected in the integration of different side information. Side information has demonstrated a high degree of effectiveness in improving recommendation performance, especially for data sparsity and cold-start issues (Zhang et al. 2016). It can appear in different forms: textual, visual, or audio information; structure or non-structure. In current studies of graph-based recommendations, side information is extensively involved either as extra attributes of nodes or edges (Song et al. 2019b), or as sources being learned to construct heterogeneous graphs (Sheu and Li 2020), or as external resources outside the graph, which are learned in parallel with the graph and then integrated at a high level. Despite the variety of utilizing side information, no study can indicate which fusion strategy is better in general cases, or suitable for which recommendation scenarios (Farseev et al. 2017). These we believe are extremely significant references and guidance for future researchers and technology users. Besides, some side information is from multi-sources such as user social relationships pointing out user–user interlinks directly (Farseev et al. 2017), while

item–item relations may be from the co-interaction pattern from a specific user or user group (Xia et al. 2021a). To integrate multi-sourced side information for recommendation purposes, some works learn representations from different graphs separately and combine the vectors from different sources (Zhang et al. 2021c). Some works combine different graphs into a large-scale heterogeneous graph which is then learned in a unified way (Wu et al. 2021a). These two kinds of integration strategies can both contribute to the improvement in recommendations, but there is no evidence showing which one is better. This is thus another research question for further study.

**Generality of Graph Learning.** Currently, no model can be applied for all types of input graphs. For existing GLRS approaches, given a data source, the common way is to take graph(s) formed by data objects and their explicit or implicit relationships as input, and the model usually needs to be reformulated and retrained under certain conditions, or a general model is applied for a specific task in a specific scenario, rather than extending the existing models for new tasks. This led to thousands of new models corresponding to thousands of different kinds of input graphs, which is far from the true generality of the graph model for recommendations. Therefore, one possible future research direction is whether there is a model suitable for learning all kinds of graphs. The dynamic property of online graphs shows the inevitability of changes in modelling input graphs. When additional attributes are added to the input data, for instance, when contextual or external information is attached to interactive data, resulting in the expansion of the graph, a model with generality should be expanded in a small range on the basis of the previously learned content to adapt to the change of the input graph rather than retraining new models. This should be one of the future research directions in the long run.

**Privacy Issue Of Graph Learning.** Though existing GLRSs reveal promising improvements on recommendation tasks, the graph topological and node binding features may cause privacy issues. The users' private information may be inferred from such a recommendation, which falls into the attribute inference attack problem. To deal with such an issue, in Zhang et al. (2021b) a privacy-preserving graph learning-based recommender is designed with a two-stage perturbation on input feature encoding and an optimization process to defend against attribute inference attacks. However, the authors also point out the challenge of balancing the personalized recommendation performance and the extent of the privacy protection mechanism, which means the privacy-preserving recommender systems are far from mature and deserve more attention in the future.

## 7 Conclusion

The study systematically investigated graph learning-based recommendation. The recommendation algorithms based on graph-structured data can be well applied to solve the sparsity and cold-start problems with improved accuracy by mining and leveraging the explicit as well as implicit relations revealed in graphs. In GLRSs, the core is how to process graph-structured data, how to learn and obtain adequate information from the graph to fulfil the final recommendation purpose, and how to adapt the graph operation process to more complex and diverse graph structures as well as large-scale node and edges in the real world. Looking at the changes of GLRSs in recent years, the graph structure has gone from homogeneous to heterogeneous, the graph attribute from zero to multiplex, the technology used from the traditional recommendation algorithm to deep learning-based models, and the evaluation of recommendation performance from focusing only on accuracy and click-through rate to increasingly multidimensional development. Such developments shed light on a new perspective for the community of recommendation researchers and practitioners. We argue that the current graph-based recommendation algorithms are far from being fully developed, and further research investment and empirical studies are still needed.

**Acknowledgements** This work is supported by the Research Council of Norway under Grant Nos. 245469 and 309834.

**Funding** Open access funding provided by NTNU Norwegian University of Science and Technology (incl St. Olavs Hospital - Trondheim University Hospital)

## Appendix A Statistics of datasets commonly used in GLRS

See Table 5.

**Table 5** Statistics of datasets commonly used in GLRS

| Domain | Dataset | Graph Type | Related Papers |
|---|---|---|---|
| **E-commerce** | Amazon (He and McAuley 2016; McAuley et al. 2015) | Tree-based Graph | He et al. (2016), Sun et al. (2017), Gao et al. (2019b), Huang et al. (2019) |
| | | Homogeneous Graph | Wang et al. (2020b), Ma et al. (2020), Zhu et al. (2021b) |
| | | K-partite Graph | Cen et al. (2019), Zheng et al. (2018), Wang et al. (2019f), Chen et al. (2020a), Wang et al. (2019d), Vijaikumar et al. (2019), Wang et al. (2019e), He et al. (2015), Sun et al. (2020), Wu et al. (2021b), Sun et al. (2021), Yang et al. (2021b), Fan et al. (2021), Zhang et al. (2022) |
| | | Complex Heterogeneous Graph | Han et al. (2018), Zhao et al. (2017), Zhao et al. (2019), Xian et al. (2019), Wang et al. (2020a), Ai et al. (2018), Ma et al. (2019), Wang et al. (2020g), Wang et al. (2021b), Xie et al. (2021), Liu et al. (2021c), Chen et al. (2021b) |
| | | Hypergraph | Wang et al. (2020c) |
| | | Multiple Graphs | Zhang et al. (2021c), Zhu et al. (2021a) |
| | | Tree-based Graph | Gao et al. (2019b) |
| | | Homogeneous Graph | Wu et al. (2019a); Song et al. (2019b); Zhu et al. (2021b) |
| | Yelp[1] | K-partite Graph | Wang et al. (2019f), Vijaikumar et al. (2019), Wang et al. (2019e), Lu et al. (2020), He et al. (2015), Yu et al. (2014), Wu et al. (2021b), Sun et al. (2021) |

**Table 5** continued

| Domain | Dataset | Graph Type | Related Papers |
|---|---|---|---|
| | | Complex Heterogeneous Graph | Zhao et al. (2017), Sun et al. (2018), Shi et al. (2018), Shi et al. (2015), Kyriakidi et al. (2020), Catherine and Cohen (2016), Lei et al. (2020a), Wang et al. (2020g), Zheng et al. (2017), Shi et al. (2016), Mezni et al. (2021), Xie et al. (2021), Liu et al. (2021c) |
| | | Hypergraph | Mao et al. (2019), Yu et al. (2021) |
| | | Multiple Graphs | Liu et al. (2020), Xia et al. (2021a), Huang et al. (2021b), Tang et al. (2021), Zhu et al. (2021a), Guo et al. (2021a) |
| | Epinions (Ma et al. 2011a; Tang et al. 2012; Massa and Avesani 2007; Zhao et al. 2014; Richardson and Domingos 2002) | Homogeneous Graph | Chen et al. (2019a), Jamali and Ester (2009), Ma et al. (2009), Ma et al. (2008), Ma et al. (2011b) |
| | | K-partite Graph | Mansoury et al. (2020) |
| | | Complex Heterogeneous Graph | Salamat et al. (2021) |
| | | Multiple Graphs | Fan et al. (2019b), Wu et al. (2019b), Liu et al. (2020), Ma et al. (2011a), Zhang et al. (2021a), Huang et al. (2021b) |
| | Diginetica[2] | Homogeneous Graph | Xu et al. (2019a), Wu et al. (2019c), Qiu et al. (2020a), Pan et al. (2020), Huang et al. (2021a) |
| | | Multiple Graphs | Wang et al. (2020h), Xia et al. (2021c) |
| | Ciao (Tang et al. 2012) | Homogeneous Graph | Chen et al. (2019a) |
| | | Complex Heterogeneous Graph | Salamat et al. (2021) |
| | | Multiple Graphs | Fan et al. (2019b), Zhang et al. (2021a) |
| | JD (Wang et al. 2015) | Tree-based Graph | Huang et al. (2019) |

**Table 5** continued

| Domain | Dataset | Graph Type | Related Papers |
|---|---|---|---|
| **E-commerce** | Retailrocket[3] | Homogeneous Graph | Xu et al. (2019a), Huang et al. (2021a) |
| | | Multiple Graphs | Xia et al. (2021c) |
| | Bbookstore (Huang et al. 2007b) | K-partite Graph | Li and Chen (2013) |
| | Clothing retail (Huang et al. 2007a) | K-partite Graph | Li and Chen (2013) |
| | Alibaba (Cen et al. 2019) | K-partite Graph | Cen et al. (2019), Tan et al. (2020), Wu et al. (2021b) |
| | | Complex Heterogeneous Graph | Wang et al. (2021b) |
| | | Multiple Graphs | Guo et al. (2021b) |
| | Taobao (Fan et al. 2019a; Zhao et al. 2019) | Homogeneous Graph | Chang et al. (2021b), Ouyang et al. (2021) |
| | | K-partite Graph | Fan et al. (2019a); Xia et al. (2021b); Chen et al. (2021a) |
| | | Complex Heterogeneous Graph | Zhao et al. (2019) |
| | | Multiple Graphs | Zhao et al. (2019), Wang et al. (2020d) |
| | Beidian (Xu et al. 2019b) | K-partite Graph | Xu et al. (2019b) |
| | YOOCHOOSE[4] | Homogeneous Graph | Wu et al. (2019c), Qiu et al. (2020a), Pan et al. (2020), Huang et al. (2021a) |
| | | Multiple Graphs | Xia et al. (2021c) |
| | Etsy (Wang et al. 2020c) | Hypergraph | Wang et al. (2020c) |
| | Beibei (Xia et al. 2021b; Gao et al. 2019a) | K-partite Graph | Xia et al. (2021b), Chen et al. (2021a) |
| | HOOPS (Fu et al. 2021) | Complex Heterogeneous Graph | Fu et al. (2021) |
| | Tmall[5] | Multiple Graphs | Wang et al. (2020h), Xia et al. (2021c), Guo et al. (2021b) |

**Table 5** continued

| Domain | Dataset | Graph Type | Related Papers |
| --- | --- | --- | --- |
| | MALib dataset (He et al. 2020) | K-partite Graph | Li et al. (2021a) |
| | Cosmetics[6] | Multiple Graphs | Liu et al. (2021b) |
| | UserBehavior[7] | Multiple Graphs | Liu et al. (2021b) |
| | Criteo[8] | Complex Heterogeneous Graph | Zheng et al. (2021) |
| | Avazu[9] | Complex Heterogeneous Graph | Zheng et al. (2021) |
| | USCFC[1]0 | Complex Heterogeneous Graph | Zheng et al. (2021) |
| | Adult[11] | Complex Heterogeneous Graph | Zheng et al. (2021) |
| | Google Play (Liang et al. 2017) | Complex Heterogeneous Graph | Xie et al. (2021) |
| | AppChina (Xie et al. 2015) | Complex Heterogeneous Graph | Xie et al. (2015) |
| | Advertising (PVC, PCC, Click) (Agarwal et al. 2010) | Tree-based Graph | Menon et al. (2011) |
| | Yahoo! shopping dataset (Kanagal et al. 2012) | Tree-based Graph | Kanagal et al. (2012) |
| | | Homogeneous Graph | Gori et al. (2007); Isufi et al. (2021); Hao et al. (2021b); Du et al. (2011); Ma et al. (2020); Ouyang et al. (2021) |

**Table 5** continued

| Domain | Dataset | Graph Type | Related Papers |
| --- | --- | --- | --- |
| **Entertainment** | MovieLens (Cantador et al. 2011; Harper and Konstan 2015) | K-partite Graph | Zheng et al. (2018), Wang et al. (2019d), Wang et al. (2020f), Zhang et al. (2019a), Lu et al. (2020), Phuong et al. (2019), Song et al. (2019a), Yu et al. (2014), Xin et al. (2019), Cheng et al. (2007), Yu et al. (2013), Jiang et al. (2018a), Nikolakopoulos and Karypis (2019), Wei et al. (2021), Musto et al. (2017), Fouss et al. (2012), Ostuni et al. (2014), Lei et al. (2020b), Tan et al. (2020), Mansoury et al. (2020), Zhang et al. (2021b), Wu et al. (2021c), Zhang et al. (2022), Fan et al. (2021), Hao et al. (2021a), Yang et al. (2021b), Hsu and Li (2021) |
| | | Complex Heterogeneous Graph | Palumbo et al. (2017), Han et al. (2018), Wang et al. (2018b), Sun et al. (2018), Wang et al. (2019a), Wang et al. (2019c), Wang et al. (2019b), Wang et al. (2019g), Yang and Dong (2020), Wang et al. (2020a), Zhang et al. (2016), Catherine and Cohen (2016), Zhou et al. (2020b), Cao et al. (2019), Shi et al. (2016), Ostuni et al. (2013), Sang et al. (2021), Lee et al. (2013), Xie et al. (2015), Palumbo et al. (2020) |
| | | Multiple Graphs | Monti et al. (2017), Wang et al. (2020d), Xia et al. (2021a), Tang et al. (2021) |
| | | Tree-based Graph | Huang et al. (2019) |
| | Last.fm (Schedl 2016; Cantador et al. 2011) | Homogeneous Graph | Chen et al. (2019a); Qiu et al. (2020b) |

**Table 5** continued

| Domain | Dataset | Graph Type | Related Papers |
|---|---|---|---|
| | | K-partite Graph | Wang et al. (2019e), Song et al. (2019a), Yao et al. (2015), Lei et al. (2020b), Wu et al. (2021c), Hao et al. (2021a) |
| | | Complex Heterogeneous Graph | Wang et al. (2019c), Wang et al. (2019b), Lei et al. (2020a), Wang et al. (2020g), Feng and Wang (2012), Ostuni et al. (2013), Sang et al. (2021), Wang et al. (2021b), Palumbo et al. (2020), Pang et al. (2022) |
| | | Hypergraph | Bu et al. (2010), Tan et al. (2011), Yu et al. (2021) |
| | | Multiple Graphs | Tian et al. (2021) |
| | YahooMusic(Monti et al. 2017; Nikolakopoulos et al. 2019; Dror et al. 2012) | Tree-based Graph | Koenigstein et al. (2011), Mnih (2012) |
| | | K-partite Graph | Nikolakopoulos and Karypis (2019) |
| | | Multiple Graphs | Monti et al. (2017), Tang et al. (2021) |
| | Flixster (Monti et al. 2017; Jamali and Ester 2010) | Homogeneous Graph | Isufi et al. (2021) |
| | | K-partite Graph | Zhang et al. (2019a) |
| | | Multiple Graphs | Monti et al. (2017), Tang et al. (2021) |
| **Entertainment** | Bing-News (Wang et al. 2018b, c) | Complex Heterogeneous Graph | Wang et al. (2018c), Wang et al. (2018b), Wang et al. (2019a), Liu et al. (2021a) |
| | KKBox's music[12] | K-partite Graph | Xin et al. (2019) |
| | | Complex Heterogeneous Graph | Wang et al. (2019g) |

**Table 5** continued

| Domain | Dataset | Graph Type | Related Papers |
|---|---|---|---|
| | HetRec Delicious (Cantador et al. 2011) | Homogeneous Graph | Song et al. (2019b) |
| | | Complex Heterogeneous Graph | Feng and Wang (2012) |
| | | Hypergraph | Zhu et al. (2016) |
| | Xing[13 14] | Homogeneous Graph | Abugabah et al. (2020) |
| | | Complex Heterogeneous Graph | Wang et al. (2020a), Pang et al. (2022) |
| | DepaulMovie (Zheng et al. 2015) | K-partite Graph | Phuong et al. (2019), Musto et al. (2021) |
| | InCarMusic (Zheng et al. 2015) | K-partite Graph | Phuong et al. (2019) |
| | Dianping-Food (Wang et al. 2019b) | Complex Heterogeneous Graph | Wang et al. (2019b) |
| | InMind Movie Agent (Catherine et al. 2017) | Complex Heterogeneous Graph | Catherine et al. (2017) |
| | IntentBooks (Zhang et al. 2016) | Complex Heterogeneous Graph | Zhang et al. (2016) |
| | IMDB (Yin et al. 2010) | Multiple Graphs | Yin et al. (2010) |
| | YouTube (Tang et al. 2009) | K-partite Graph | Cen et al. (2019) |
| | Adressa (Gulla et al. 2017) | Complex Heterogeneous Graph | Sheu and Li (2020), Shi et al. (2021) |
| | | Hypergraph | Gharahighehi et al. (2021) |
| | Roularta (Gharahighehi et al. 2021) | Hypergraph | Gharahighehi et al. (2021), Gharahighehi et al. (2020) |
| | MIND[15] | Complex Heterogeneous Graph | Wu et al. (2021a), Liu et al. (2021a), Zhang et al. (2021d) |
| | Multiple News Portal (Li et al. 2011) | Hypergraph | Li and Li (2013) |

**Table 5** continued

| Domain | Dataset | Graph Type | Related Papers |
|---|---|---|---|
| | Sougou News (Shi et al. 2021) | Complex Heterogeneous Graph | Shi et al. (2021) |
| | Filmtrust (Guo et al. 2013) | Multiple Graphs | Zhang et al. (2021a) |
| | Mtime (Li et al. 2014) | Homogeneous Graph | Li et al. (2014) |
| | Netease (Cao et al. 2017) | Multiple Graphs | Chang et al. (2021a) |
| | Kuaishou (Chang et al. 2021b) | Homogeneous Graph | Chang et al. (2021b) |
| | Tiktok (Wei et al. 2021) | K-partite Graph | Wei et al. (2021) |
| | Kwai (Wei et al. 2021) | K-partite Graph | Wei et al. (2021) |
| | Restaurant&consumer (Vargas-Govea et al. 2011) | K-partite Graph | Jiang et al. (2018a) |
| **Social Network** | Douban (Song et al. 2019b; Ma et al. 2011a; Zheng et al. 2017; Shi et al. 2018; Monti et al. 2017) | Homogeneous Graph | Song et al. (2019b), Isufi et al. (2021) |
| | | K-partite Graph | Zhang et al. (2019a) |
| | | Complex Heterogeneous Graph | Shi et al. (2018), Shi et al. (2015), Salamat et al. (2021) |
| | | Hypergraph | Yu et al. (2021) |
| | | Multiple Graphs | Ma et al. (2011a), Zheng et al. (2017), Shi et al. (2016), Monti et al. (2017), Zhang et al. (2021a), Tian et al. (2021), Guo et al. (2021a) |
| | WeChat (Wu et al. 2019b; Wang et al. 2020f) | K-partite Graph | Wang et al. (2020f) |
| | | Multiple Graphs | Wu et al. (2019b) |

**Table 5** continued

| Domain | Dataset | Graph Type | Related Papers |
|---|---|---|---|
| | Twitter (Sharma et al. 2016; De Domenico et al. 2013) | K-partite Graph | Cen et al. (2019); Sharma et al. (2016) |
| | Reddit[16] | Homogeneous Graph | Abugabah et al. (2020) |
| | | Complex Heterogeneous Graph | Pang et al. (2022) |
| | Pinterest(Ying et al. 2018; Eksombatchai et al. 2018) | K-partite Graph | Ying et al. (2018); Eksombatchai et al. (2018); Lei et al. (2020b); Tan et al. (2020); Yang et al. (2021b) |
| | Flickr (Wu et al. 2019a) | Homogeneous Graph | Wu et al. (2019a) |
| | Hike network (Verma et al. 2019) | Multiple Graphs | Verma et al. (2019) |
| **Academic or Book** | BookCrossing (Ziegler et al. 2005) | K-partite Graph | Fouss et al. (2012); Li and Chen (2013); Hsu and Li (2021) |
| | | Complex Heterogeneous Graph | Wang et al. (2018b), Wang et al. (2019a), Wang et al. (2019c), Wang et al. (2019b), Yang and Dong (2020), Zhou et al. (2020b), Sang et al. (2021) |
| | DBbook[17] (Lu et al. 2020) | K-partite Graph | Song et al. (2019a); Musto et al. (2017); Lu et al. (2020) |
| | | Complex Heterogeneous Graph | Cao et al. (2019) |
| | DBLP (Tang et al. 2008) | Complex Heterogeneous Graph | Zhu et al. (2021c) |
| | | Multiple Graphs | Ali et al. (2020); Yin et al. (2010) |
| | CiteULike (Wang et al. 2013) | Homogeneous Graph | Gao et al. (2018) |
| | | K-partite Graph | Yao et al. (2015); Song et al. (2011) |
| | | Hypergraph | Zhu et al. (2016) |

**Table 5** continued

| Domain | Dataset | Graph Type | Related Papers |
|---|---|---|---|
| | Goodreads (Wang and Caverlee 2019) | Homogeneous Graph | Ma et al. (2020) |
| | | Hypergraph | Wang et al. (2020c) |
| | User-Tag (Pan et al. 2008; Sun et al. 2015; Song et al. 2011) | Homogeneous Graph | Du et al. (2011); Sun et al. (2015) |
| | | K-partite Graph | Song et al. (2011) |
| | Librarything (Zhao et al. 2014) | Complex Heterogeneous Graph | Palumbo et al. (2020) |
| | | Multiple Graphs | Liu et al. (2020) |
| | Youshu (Chen et al. 2019b) | Multiple Graphs | Chang et al. (2021a) |
| | ACL Anthology Network (Radev et al. 2013) | Multiple Graphs | Ali et al. (2020) |
| | OHSUMED (Hersh et al. 1994) | K-partite Graph | Cheng et al. (2007) |
| | | K-partite Graph | Wang et al. (2019f); Chen et al. (2020a); Tan et al. (2020) |

**Table 5** continued

| Domain | Dataset | Graph Type | Related Papers |
|---|---|---|---|
| | Gowalla (Liang et al. 2016; Cho et al. 2011; Liu et al. 2013) | Homogeneous Graph | Qiu et al. (2020b); Hao et al. (2021b) |
| | | Complex Heterogeneous Graph | Bagci and Karagoz (2016) |
| | | Multiple Graphs | Christoforidis et al. (2021) |
| **POI** | Trip.com+Facebook+Twitter (Wang et al. 2017) | Multiple Graphs | Wang et al. (2017) |
| | Foursquare (Gao et al. 2012) | Tree-based Graph | Yang et al. (2016) |
| | | K-partite Graph | Wang et al. (2019d) |
| | | Complex Heterogeneous Graph | Bagci and Karagoz (2016) |
| | | Multiple Graphs | Christoforidis et al. (2021) |
| | Brightkite (Cho et al. 2011) | Complex Heterogeneous Graph | Bagci and Karagoz (2016) |
| | Google Local (He et al. 2017) | Homogeneous Graph | Zhu et al. (2021b) |
| | | Multiple Graphs | Zhu et al. (2021a) |
| | Tripadvisor[18] | K-partite Graph | Musto et al. (2021) |
| | NUS-MSS (Farseev et al. 2015) | Homogeneous Graph | Farseev et al. (2017) |
| **Others** | Educational website and textbooks (Shi et al. 2020) | Complex Heterogeneous Graph | Shi et al. (2020) |
| | MIT AI + CASAS (Chen et al. 2019d) | Complex Heterogeneous Graph | Chen et al. (2019d) |
| | REDIAL (Li et al. 2018b) | Complex Heterogeneous Graph | Zhou et al. (2020a) |
| | VizML corpus (Hu et al. 2019) | K-partite Graph | Li et al. (2021b) |

**Table 5** continued

| Domain | Dataset | Graph Type | Related Papers |
|---|---|---|---|
| | Legal recommendation dataset (Yang et al. 2021a) | Complex Heterogeneous Graph | Yang et al. (2021a) |
| | Yahoo! traffic stream (Menon et al. 2011) | Tree-based Graph | Menon et al. (2011) |

[1] https://www.yelp.com/dataset/documentation/main
[2] https://competitions.codalab.org/competitions/11161
[3] https://www.kaggle.com/retailrocket/ecommerce-dataset
[4] https://2015.recsyschallenge.com/challenge.html
[5] https://tianchi.aliyun.com/dataset/dataDetail?dataId=42
[6] https://www.kaggle.com/mkechinov/ecommerce-events-history-in-cosmetics-shop
[7] https://tianchi.aliyun.com/dataset/dataDetail?dataId=649&userId=1
[8] https://www.kaggle.com/c/criteo-display-ad-challenge
[9] https://www.kaggle.com/c/avazu-ctr-prediction
[10] https://www.kaggle.com/kaggle/us-consumer-finance-complaints
[11] http://archive.ics.uci.edu/ml/datasets/Adult
[12] https://wsdm-cup-2018.kkbox.events/
[13] https://2016.recsyschallenge.com/
[14] http://www.recsyschallenge.com/2017/
[15] https://msnews.github.io/
[16] https://www.kaggle.com/colemaclean/subreddit-interactions
[17] http://2014.eswc-conferences.org/important-dates/call-RecSys.html
[18] https://github.com/irecsys/CARSKit/tree/master/context-aware_data_sets

## Appendix B Technology summary in GLRS

See Table 6.

**Table 6** Technology summary in GLRS

| Technology Class | Tech Subclass | Domain | Dataset | Graph Type |
|---|---|---|---|---|
| Traditional Machine Learning Techniques | PRM | E-commerce | Yelp | Complex Heterogeneous Graph: Catherine and Cohen (2016) |
| | | Entertainment | InMind Movie Agent | Complex Heterogeneous Graph: Catherine et al. (2017) |
| | | | MovieLens | Complex Heterogeneous Graph: Catherine and Cohen (2016), Lee et al. (2013) |
| | | | | K-partite Graph: Shams and Haratizadeh (2017), Musto et al. (2017) |
| | | | Last.fm | K-partite Graph: Jäschke et al. (2007) |
| | | | DePaulMovie | K-partite Graph: Musto et al. (2021) |
| | | Academic or Book | BibSonomy | K-partite Graph: Jäschke et al. (2007) |
| | | | DBbook | K-partite Graph: Musto et al. (2017) |
| | | POI | Tripadvisor | K-partite Graph: Musto et al. (2021) |
| | RTGR | E-commerce | Yelp | Hypergraph: Mao et al. (2019), Yu et al. (2021) |
| | | Entertainment | Last.fm | Hypergraph: Bu et al. (2010), Tan et al. (2011), Yu et al. (2021) |
| | | | Multiple News Portal | Hypergraph: Li and Li (2013) |
| | | | HetRec Delicious | Hypergraph: Zhu et al. (2016) |
| | | | Adressa | Hypergraph: Gharahighehi et al. (2021) |
| | | | Roularta | Hypergraph: Gharahighehi et al. (2021) |
| | | Social Network | Douban | Hypergraph: Yu et al. (2021) |
| | | Academic or Book | CiteULike | Hypergraph: Zhu et al. (2016) |

**Table 6** continued

| Technology Class | Tech Subclass | Domain | Dataset | Graph Type |
|---|---|---|---|---|
| | | POI | Trip.com+Facebook+Twitter | Multiple Graphs: Gao et al. (2019b) |
| | KML | Entertainment | MovieLens | K-partite Graph: Fouss et al. (2012) |
| | | | | Complex Heterogeneous Graph: Ostuni et al. (2014) |
| | | | mtime | Homogeneous Graph: Li et al. (2014) |
| | | Academic or Book | Book-Crossing | K-partite Graph: Fouss et al. (2012) |
| | LFM | E-commerce | Amazon | Complex Heterogeneous Graph: Zhao et al. (2017) |
| | | | | Tree-based Graph: Sun et al. (2017) |
| | | | Yelp | Complex Heterogeneous Graph: Zhao et al. (2017), Kyriakidi et al. (2020) |
| | | | Epinions | Multiple Graphs: Ma et al. (2011a) |
| | | | | Homogeneous Graph: Ma et al. (2009), Ma et al. (2008), Ma et al. (2011b) |
| | | | Yahoo! Shopping | Tree-based Graph: Kanagal et al. (2012) |
| | LFM | Entertainment | Yahoo!Music | Tree-based Graph: Koenigstein et al. (2011), Mnih (2012) |
| | | | MovieLens | Complex Heterogeneous Graph: Du et al. (2011) |
| | | Social Network | Douban | Multiple Graphs: Ma et al. (2011a) |
| | | Academic or Book | User-Tag | Homogeneous Graph: Du et al. (2011), Sun et al. (2015) |
| | | POI | Foursquare | Tree-based Graph: Yang et al. (2016) |
| | | Other | Yahoo! traffic stream | Tree-based Graph: Menon et al. (2011) |
| Traditional Machine Learning Techniques | Others (Spectral graph) | E-commerce | Amazon | K-partite Graph: Zheng et al. (2018) |
| | | Entertainment | MovieLens | K-partite Graph: Zheng et al. (2018) |
| | | | HetRec Delicious | K-partite Graph: Zheng et al. (2018) |

**Table 6** continued

| Technology Class | Tech Subclass | Domain | Dataset | Graph Type |
|---|---|---|---|---|
| | | POI | NUS-MSS | Homogeneous Graph: Farseev et al. (2017) |
| | Others (Similarity) | Entertainment | DepaulMovie | K-partite Graph: Phuong et al. (2019) |
| | | | MovieLens | K-partite Graph: Phuong et al. (2019) |
| | | | InCarMusic | K-partite Graph: Phuong et al. (2019) |
| | Others (Clustering) | Social Network | User-Tag | K-partite Graph: Song et al. (2011) |
| | | | BbSonomy | K-partite Graph: Song et al. (2011) |
| | | Academic or Book | CiteULike | K-partite Graph: Song et al. (2011) |
| | Others (Complex Number) | E-commerce | AppChina | Complex Heterogeneous Graph: Xie et al. (2015) |
| | | Entertainment | MovieLens | Complex Heterogeneous Graph: Xie et al. (2015) |
| | | E-commerce | Clothing-Retail | K-partite Graph: Li and Chen (2013) |
| | | | Amazon | Multiple Graphs: Vijaikumar et al. (2019) |
| | | | | Complex Heterogeneous Graph: Ma et al. (2019) |
| Path-based Techniques | RWM | | Yelp | Multiple Graphs: Vijaikumar et al. (2019) |
| | | | Epinions | Homogeneous Graph: Jamali and Ester (2009) |
| | | Entertainment | IMDB | Multiple Graphs: Yin et al. (2010) |
| | | | MovieLens | K-partite Graph: Cheng et al. (2007), Jiang et al. (2018a), Nikolakopoulos and Karypis (2019) |
| | | | | Homogeneous Graph: Gori et al. (2007) |
| | | | Last.fm | K-partite Graph: Yao et al. (2015) |
| | | | | Complex Heterogeneous Graph: Feng and Wang (2012) |
| | | | Yahoo2RMusic | K-partite Graph: Nikolakopoulos and Karypis (2019) |

**Table 6** continued

| Technology Class | Tech Subclass | Domain | Dataset | Graph Type |
|---|---|---|---|---|
| | | | HetRec Delicious | Complex Heterogeneous Graph: Feng and Wang (2012) |
| | | | Restaurant & Consumer | K-partite Graph: Jiang et al. (2018a) |
| | | Social Network | Twitter | K-partite Graph: Sharma et al. (2016) |
| | | | Pinterest | K-partite Graph: Eksombatchai et al. (2018) |
| Path-based Techniques | RWM | Academic or Book | Book-Crossing | K-partite Graph: Li and Chen (2013) |
| | | | DBLP | Multiple Graphs: Yin et al. (2010) |
| | | | OHSUMED | K-partite Graph: Cheng et al. (2007) |
| | | | CiteULike | K-partite Graph: Yao et al. (2015) |
| | | POI | Brightkite | Complex Heterogeneous Graph: Bagci and Karagoz (2016) |
| | | | Gowalla | Complex Heterogeneous Graph: Bagci and Karagoz (2016) |
| | | | Foursquare | Complex Heterogeneous Graph: Bagci and Karagoz (2016) |
| | RM | E-commerce | Amazon | K-partite Graph: He et al. (2015) |
| | | | Yelp | K-partite Graph: He et al. (2015) |
| | | Academic or Book | Educational Website and Textbooks | Complex Heterogeneous Graph: Shi et al. (2020) |
| | MPM | E-commerce | Amazon | K-partite Graph: Cen et al. (2019) |
| | | | Alibaba | K-partite Graph: Cen et al. (2019) |
| | | | Yelp | K-partite Graph: Lu et al. (2020); Yu et al. (2014) |
| | | | | Complex Heterogeneous Graph: Shi et al. (2018), Shi et al. (2015), Zheng et al. (2017), Shi et al. (2016) |

**Table 6** continued

| Technology Class | Tech Subclass | Domain | Dataset | Graph Type |
|---|---|---|---|---|
| | | Entertainment | YouTube | K-partite Graph: Cen et al. (2019) |
| | | | MovieLens | K-partite Graph: Lu et al. (2020), Yu et al. (2014), Yu et al. (2013) |
| | | | | Complex Heterogeneous Graph: Shi et al. (2016), Ostuni et al. (2013) |
| | | | Last.fm | Complex Heterogeneous Graph: Ostuni et al. (2013) |
| | | Social Network | Twitter | K-partite Graph: Cen et al. (2019) |
| | | | Douban | Complex Heterogeneous Graph: Shi et al. (2018), Shi et al. (2015), Zheng et al. (2017), Shi et al. (2016) |
| | | Academic or Book | DBook | K-partite Graph: Lu et al. (2020) |
| Graph Embedding-based Techniques | GDRM | E-commerce | Amazon | Complex Heterogeneous Graph: Fu et al. (2020), Wang et al. (2020a) |
| | | | | Tree-based Graph: He et al. (2016) |
| | | | Taobao | Multiple Graphs: Wang et al. (2020d) |
| | | Entertainment | MovieLens | Complex Heterogeneous Graph: Wang et al. (2020a); Palumbo et al. (2020) |
| | | | | Multiple Graphs: Wang et al. (2020d) |
| | | | Last.fm | Complex Heterogeneous Graph: Palumbo et al. (2020) |
| | | | Xing | Complex Heterogeneous Graph: Wang et al. (2020a) |
| | GDRM | Social Network | Hike Network | Multiple Graphs: Verma et al. (2019) |
| | | Academic or Book | Citation | Complex Heterogeneous Graph: Jiang et al. (2018b) |
| | | | LibraryThing | Complex Heterogeneous Graph: Palumbo et al. (2020) |

**Table 6** continued

| Technology Class | Tech Subclass | Domain | Dataset | Graph Type |
|---|---|---|---|---|
| | | | DBLP | Multiple Graphs: Ali et al. (2020) |
| | | | ACL Anthology Network | Multiple Graphs: Ali et al. (2020) |
| | | POI | Foursquare | Multiple Graphs: Christoforidis et al. (2021) |
| Graph Embedding-based Techniques | | | Gowalla | Multiple Graphs: Christoforidis et al. (2021) |
| | | E-commerce | Amazon | Complex Heterogeneous Graph: Ai et al. (2018) |
| | | | | Homogeneous Graph: Wang et al. (2020b) |
| | TEM | | Alibaba | Complex Heterogeneous Graph: Wang et al. (2021b) |
| | | Entertainment | MovieLens | Complex Heterogeneous Graph: Palumbo et al. (2017), Wang et al. (2018b), Cao et al. (2019) |
| | | | Bing-News | Complex Heterogeneous Graph: Wang et al. (2018b) |
| | | | Last.fm | Complex Heterogeneous Graph: Wang et al. (2021b) |
| | | | Book-Crossing | Complex Heterogeneous Graph: Wang et al. (2018b) |
| | | Academic or Book | CiteULike | Homogeneous Graph: Gao et al. (2018) |
| | | | DBBook | Complex Heterogeneous Graph: Cao et al. (2019) |
| | | Other | MIT AI+CASAS | Complex Heterogeneous Graph: Chen et al. (2019d) |
| | | | VizML Corpus | K-partite Graph: Li et al. (2021b) |
| | | E-commerce | Amazon | Tree-based Graph: Huang et al. (2019) |
| | | | JD | Tree-based Graph: Huang et al. (2019) |
| | DNN | | Yelp | Homogeneous Graph: Wu et al. (2019a) |

**Table 6** continued

| Technology Class | Tech Subclass | Domain | Dataset | Graph Type |
|---|---|---|---|---|
| | | | | Complex Heterogeneous Graph: Sun et al. (2018); Mezni et al. (2021) |
| | | Entertainment | MovieLens | Complex Heterogeneous Graph: Sun et al. (2018), Wang et al. (2019g) |
| | | | Last.fm | Tree-based Graph: Huang et al. (2019) |
| | | | KKBox | Complex Heterogeneous Graph: Wang et al. (2019g) |
| | | Social Network | Flickr | Homogeneous Graph: Wu et al. (2019a) |
| | AE | Entertainment | MovieLens | Complex Heterogeneous Graph: Zhang et al. (2016) |
| | | Academic or Book | IntentBooks | Complex Heterogeneous Graph: Zhang et al. (2016) |
| | | | Amazon | Tree-based Graph: Gao et al. (2019b) |
| | | | | K-partite Graph: Wang et al. (2019d); Zhang et al. (2022) |
| | | | | Complex Heterogeneous Graph: Han et al. (2018) |
| | | | Yelp | Tree-based Graph: Gao et al. (2019b) |
| | | | | Multiple Graphs: Xia et al. (2021a) |
| Deep Learning-based Techniques | AM | E-commerce | Epinions | Homogeneous Graph: Chen et al. (2019a) |
| | | | Ciao | Homogeneous Graph: Chen et al. (2019a) |
| | | | HOOPS | Complex Heterogeneous Graph: Fu et al. (2021) |
| | | | Taobao | Homogeneous Graph: Chang et al. (2021b) |
| | | | Kuaishou | Homogeneous Graph: Chang et al. (2021b) |
| | | Entertainment | MovieLens | Complex Heterogeneous Graph: Han et al. (2018), Wang et al. (2019a) |
| | | | | Homogeneous Graph: Hao et al. (2021b) |

**Table 6** continued

| Technology Class | Tech Subclass | Domain | Dataset | Graph Type |
|---|---|---|---|---|
| | | | | K-partite Graph: Wang et al. (2019d); Xin et al. (2019); Zhang et al. (2022) |
| | | | | Multiple Graphs: Xia et al. (2021a) |
| | | | Bing-News | Complex Heterogeneous Graph: Wang et al. (2019a) |
| | | | KKBox | K-partite Graph: Xin et al. (2019) |
| | | | Last.fm | Homogeneous Graph: Chen et al. (2019a) |
| | | Academic or Book | Book-Crossing | Complex Heterogeneous Graph: Wang et al. (2019a) |
| | | POI | Foursquare | K-partite Graph: Wang et al. (2019d) |
| | | | Gowalla | Homogeneous Graph: Hao et al. (2021b) |
| | DRL | E-commerce | Amazon | Complex Heterogeneous Graph: Xian et al. (2019) |
| | | | Yelp | Complex Heterogeneous Graph: Lei et al. (2020a) |
| | | Entertainment | Last.fm | K-partite Graph: Song et al. (2019a) |
| | | | | Complex Heterogeneous Graph: Lei et al. (2020a) |
| | DRL | | MovieLens | K-partite Graph: Song et al. (2019a) |
| | | | MIND | Complex Heterogeneous Graph: Liu et al. (2021a) |
| | | | Bing-News | Complex Heterogeneous Graph: Liu et al. (2021a) |
| | | Academic or Book | DBbook | K-partite Graph: Song et al. (2019a) |
| | | | Amazon | K-partite Graph: Wang et al. (2019f); Chen et al. (2020a); Wang et al. (2019e); Sun et al. (2020); Wu et al. (2021b) |
| | | | | Homogeneous Graph: Zhu et al. (2021b), Ma et al. (2020) |

**Table 6** continued

| Technology Class | Tech Subclass | Domain | Dataset | Graph Type |
|---|---|---|---|---|
| Deep Learning-based Techniques | GNN | E-commerce | | Complex Heterogeneous Graph: Zhao et al. (2019) |
| | | | | Multiple Graphs: Zhang et al. (2021c); Zhu et al. (2021a) |
| | | | Yelp | Homogeneous Graph: Zhu et al. (2021b) |
| | | | | K-partite Graph: Wang et al. (2019f), Wang et al. (2019e), Wu et al. (2021b), Sun et al. (2021), Sun et al. (2021), Yang et al. (2021b) |
| | | | | Multiple Graphs: Liu et al. (2020), Huang et al. (2021b), Tang et al. (2021), Zhu et al. (2021a), Guo et al. (2021a) |
| | | | Epinions | Multiple Graphs: Wu et al. (2019b); Liu et al. (2020); Huang et al. (2021b) |
| | | | Taobao | Homogeneous Graph: Ouyang et al. (2021) |
| | | | | K-partite Graph: Fan et al. (2019a), Chen et al. (2021a) |
| | | | | Complex Heterogeneous Graph: Zhao et al. (2019) |
| | | | Alibaba | K-partite Graph: Tan et al. (2020), Wu et al. (2021b) |
| | | | | Multiple Graphs: Guo et al. (2021b) |
| | | | Beibei | K-partite Graph: Chen et al. (2021a) |
| | | | Criteo | Complex Heterogeneous Graph: Zheng et al. (2021) |
| | | | Tmall | Multiple Graphs: Wang et al. (2020h), Guo et al. (2021b) |
| | | | Diginetica | Homogeneous Graph: Qiu et al. (2020a), Pan et al. (2020) |
| | | | | Multiple Graphs: Wang et al. (2020h), Xia et al. (2021c) |

**Table 6** continued

| Technology Class | Tech Subclass | Domain | Dataset | Graph Type |
|---|---|---|---|---|
| | | | Yoochoose | Homogeneous Graph: Qiu et al. (2020a), Pan et al. (2020) |
| | | | | Multiple Graphs: Xia et al. (2021c) |
| | | | Avazu | Complex Heterogeneous Graph: Zheng et al. (2021) |
| | | | USCFC | Complex Heterogeneous Graph: Zheng et al. (2021) |
| | | | Adult | Complex Heterogeneous Graph: Zheng et al. (2021) |
| | | Entertainment | MovieLens | Homogeneous Graph: Ma et al. (2020), Isufi et al. (2021), Ouyang et al. (2021) |
| | | | | K-partite Graph: Lei et al. (2020b), Tan et al. (2020), Yang et al. (2021b), Zhang et al. (2021b) |
| | | | | Complex Heterogeneous Graph: Wang et al. (2019c), Wang et al. (2019b) |
| | | | | Multiple Graph: Tang et al. (2021) |
| Deep Learning-based Techniques | GNN | | Last.fm | Homogeneous Graph: Qiu et al. (2020b) |
| | | | | K-partite Graph: Wang et al. (2019e), Lei et al. (2020b) |
| | | | | Complex Heterogeneous Graph: Wang et al. (2019c), Wang et al. (2019b) |
| | | | | Multiple Graphs: Tian et al. (2021) |
| | | | Dianping-Food | Complex Heterogeneous Graph: Wang et al. (2019b) |
| | | | Flixster | Homogeneous Graph: Isufi et al. (2021) |
| | | | | Multiple Graphs: Tang et al. (2021) |

**Table 6** continued

| Technology Class | Tech Subclass | Domain | Dataset | Graph Type |
|---|---|---|---|---|
| | | | MIND | Complex Heterogeneous Graph: Wu et al. (2021a), Zhang et al. (2021d) |
| | | | Netease | Multiple Graphs: Chang et al. (2021a) |
| | | | YahooMusic | Multiple Graphs: Tang et al. (2021) |
| | | | Tiktok | K-partite Graph: Wei et al. (2021) |
| | | | Kwai | K-partite Graph: Wei et al. (2021) |
| | | Social Network | WeChat | Multiple Graphs: Wu et al. (2019b) |
| | | | Pinterest | K-partite Graph: Ying et al. (2018), Lei et al. (2020b), Tan et al. (2020), Yang et al. (2021b) |
| | | | Douban | Homogeneous Graph: Isufi et al. (2021) |
| | | | | Multiple Graphs: Tian et al. (2021), Guo et al. (2021a) |
| | | Academic or Book | Book-Crossing | Complex Heterogeneous Graph: Wang et al. (2019c), Wang et al. (2019b) |
| | | | Librarything | Multiple Graphs: Liu et al. (2020) |
| | | | Youshu | Multiple Graphs: Chang et al. (2021a) |
| | | | Goodreads | Homogeneous Graph: Ma et al. (2020) |
| | | POI | Gowalla | Homogeneous Graph: Qiu et al. (2020b) |
| | | | | K-partite Graph: Wang et al. (2019f), Chen et al. (2020a), Tan et al. (2020) |
| | | | Google Local | Homogeneous Graph: Zhu et al. (2021b) |
| | | | | Multiple Graphs: Zhu et al. (2021a) |

**Table 6** continued

| Technology Class | Tech Subclass | Domain | Dataset | Graph Type |
|---|---|---|---|---|
| | | Other | MALib-Dataset | K-partite Graph: Li et al. (2021a) |
| | | | Amazon | K-partite Graph: Fan et al. (2021) |
| | | | | Complex Heterogeneous Graph: Wang et al. (2020g), Liu et al. (2021c), Xie et al. (2021), Chen et al. (2021b) |
| | | | | Hypergraph: Wang et al. (2020c) |
| | | | Yelp | Homogeneous Graph: Song et al. (2019b) |
| | | | | Complex Heterogeneous Graph: Wang et al. (2020g), Liu et al. (2021c), Xie et al. (2021) |
| Deep Learning-based Techniques | DHM | E-commerce | Diginetica | Homogeneous Graph: Xu et al. (2019a); Wu et al. (2019c); Huang et al. (2021a) |
| | | | | Multiple Graphs: Xia et al. (2021c) |
| | | | Retailrocket | Homogeneous Graph: Xu et al. (2019a), Huang et al. (2021a) |
| | | | Yoochoose | Homogeneous Graph: Wu et al. (2019c), Huang et al. (2021a) |
| | | | Ciao | Complex Heterogeneous Graph: Salamat et al. (2021) |
| | | | | Multiple Graphs: Fan et al. (2019b), Zhang et al. (2021a) |
| | | | Epinions | Complex Heterogeneous Graph: Salamat et al. (2021) |
| | | | | Multiple Graphs: Fan et al. (2019b); Zhang et al. (2021a) |
| | | | Etsy | Hypergraph: Wang et al. (2020c) |

**Table 6** continued

| Technology Class | Tech Subclass | Domain | Dataset | Graph Type |
|---|---|---|---|---|
| | | | Beidia | K-partite Graph: Xu et al. (2019b) |
| | | | Taobao | K-partite Graph: Xia et al. (2021b) |
| | | | Beibei | K-partite Graph: Xia et al. (2021b) |
| | | | Tmall | Multiple Graphs: Xia et al. (2021c) |
| | | | RetailRocket | Multiple Graphs: Xia et al. (2021c) |
| | | | Cosmetics | Multiple Graphs: Liu et al. (2021b) |
| | | | UserBehavior | Multiple Graphs: Liu et al. (2021b) |
| | | | Google Play | Complex Heterogeneous Graph: Xie et al. (2021) |
| | | Entertainment | MovieLens | K-partite Graph: Wang et al. (2020f), Zhang et al. (2019a), Wu et al. (2021c), Fan et al. (2021), Hao et al. (2021a), Hsu and Li (2021) |
| | | | | Complex Heterogeneous Graph: Yang and Dong (2020), Zhou et al. (2020b), Sang et al. (2021) |
| | | | | Multiple Graphs: Monti et al. (2017) |
| Deep Learning-based Techniques | DHM | Entertainment | Last.fm | K-partite Graph: Wu et al. (2021c), Hao et al. (2021a) |
| | | | | Complex Heterogeneous Graph: Yang and Dong (2020), Wang et al. (2020g), Sang et al. (2021), Pang et al. (2022) |

**Table 6** continued

| Technology Class | Tech Subclass | Domain | Dataset | Graph Type |
|---|---|---|---|---|
| | | Entertainment | HetRec Delicious | Homogeneous Graph: Song et al. (2019b) |
| | | | Adressa | Complex Heterogeneous Graph: Sheu and Li (2020), Shi et al. (2021) |
| | | | Roularta | Hypergraph: Gharahighehi et al. (2020) |
| | | | Bing-News | Complex Heterogeneous Graph: Wang et al. (2018c) |
| | | | Flixster | K-partite Graph: Zhang et al. (2019a) |
| | | | | Multiple Graphs: Monti et al. (2017) |
| | | | Xing | Homogeneous Graph: Abugabah et al. (2020) |
| | | | | Complex Heterogeneous Graph: Pang et al. (2022) |
| Deep Learning-based Techniques | DHM | | YahooMusic | Multiple Graphs: Monti et al. (2017) |
| | | | Filmtrust | Multiple Graphs: Zhang et al. (2021a) |
| | | | Sougou | Complex Heterogeneous Graph: Shi et al. (2021) |
| | | | Douban | Homogeneous Graph: Song et al. (2019b) |
| | | Social Network | | K-partite Graph: Zhang et al. (2019a) |
| | | | | Complex Heterogeneous Graph: Salamat et al. (2021) |
| | | | | Multiple Graphs: Monti et al. (2017), Zhang et al. (2021a) |

**Table 6** continued

| Technology Class | Tech Subclass | Domain | Dataset | Graph Type |
|---|---|---|---|---|
| | | | WeChat | K-partite Graph: Wang et al. (2020f) |
| | | | Reddit | Homogeneous Graph.: Abugabah et al. (2020) |
| | | | | Complex Heterogeneous Graph: Pang et al. (2022) |
| | | | Others | K-partite Graph: Kim et al. (2019) |
| | | Academic or Book | Goodreads | Hypergraph: Wang et al. (2020c) |
| | | | Book-Crossing | K-partite Graph: Hsu and Li (2021) |
| | | | | Complex Heterogeneous Graph: Yang and Dong (2020), Zhou et al. (2020b), Sang et al. (2021) |
| | | | DBLP | Complex Heterogeneous Graph: Zhu et al. (2021c) |
| | | Others | REDIAL | Complex Heterogeneous Graph: Zhou et al. (2020a) |
| | | | Legal recommendation dataset | Complex Heterogeneous Graph: Yang et al. (2021a) |

## Appendix C Statistics of datasets and related technologies in GLRS

See Table 7.

**Table 7** Quantitative summarization of different datasets on adopted GLRS technologies

| Domain | Dataset | Traditional ML | Path-based | Graph embedding | Deep Learning | | | | | |
|---|---|---|---|---|---|---|---|---|---|---|
| | | | | | DNN | AE | AM | DRL | GNN | DHM |
| **E-commerce** | Amazon | 3 | 4 | 5 | 1 | | 4 | 1 | 12 | 6 |
| | Yelp | 5 | 8 | | 3 | | 2 | 1 | 9 | 4 |
| | Epinions | 4 | 1 | | | | 1 | | 3 | 3 |
| | Diginetica | | | | | | | | 4 | 4 |
| | Ciao | | | | | | 1 | | | 3 |
| | JD | | | | 1 | | | | | |
| | Retailrocket | | | | | | | | | 3 |
| | Clothing retail | | 1 | | | | | | | |
| | Alibaba | | 1 | 1 | | | | | 3 | |
| | Taobao | | | 1 | | | 1 | | 4 | 1 |
| | Beidian | | | | | | | | | |
| | YOOCHOOSE | | | | | | | | 3 | 2 |
| | Etsy | | | | | | | | | 1 |
| | Beibei | | | | | | | | 1 | 1 |
| | HOOPS | | | | | | 1 | | | |
| | Tmall | | | | | | | | 2 | 1 |
| | Cosmetics | | | | | | | | | 1 |
| | UserBehavior | | | | | | | | | 1 |
| | Criteo | | | | | | | | 1 | |
| | Avazu | | | | | | | | 1 | |
| | USCFC | | | | | | | | 1 | |
| | Adult | | | | | | | | 1 | |
| | Google Play | | | | | | | | | 1 |
| | AppChina | 1 | | | | | | | | |
| | Yahoo! shopping dataset | 1 | | | | | | | | |
| | MovieLens | 10 | 9 | 6 | 2 | 1 | 7 | 1 | 9 | 10 |
| | Last.fm | 4 | 3 | 2 | 1 | | 1 | 2 | 6 | 6 |
| | YahooMusic | 2 | 1 | | | | | | 1 | 1 |
| | Flixster | | | | | | | | 2 | 2 |
| | Bing-News | | | 1 | | | 1 | 1 | | 1 |
| | KKBox's music | | | | 1 | | 1 | | | |

**Table 7** continued

| Domain | Dataset | Traditional ML | Path-based | Graph embedding | Deep Learning | | | | | |
|---|---|---|---|---|---|---|---|---|---|---|
| | | | | | DNN | AE | AM | DRL | GNN | DHM |
| **Entertainment** | HetRec Delicious | 2 | 1 | | | | | | | 1 |
| | Xing | | | 1 | | | | | | 2 |
| | DepaulMovie | 2 | | | | | | | | |
| | InCarMusic | 1 | | | | | | | | |
| | Dianping-Food | | | | | | | | 1 | |
| | InMind Movie Agent | 1 | | | | | | | | |
| | IntentBooks | | | | | 1 | | | | |
| | IMDB | | 1 | | | | | | | |
| | YouTube | | 1 | | | | | | | |
| | Adressa | 2 | | | | | | | | 2 |
| | Roularta | 2 | | | | | | | | 1 |
| | MIND | | | | | | | 1 | 2 | |
| | Multiple News Portal | 1 | | | | | | | | |
| | Sougou News | | | | | | | | | 1 |
| | Filmtrust | | | | | | | | | 1 |
| | Mtime | 1 | | | | | | | | |
| | Netease | | | | | | | | 1 | |
| | Kuaishou | | | | | | 1 | | | |
| | Tiktok | | | | | | | | 1 | |
| | Kwai | | | | | | | | 1 | |
| | Restaurant &consumer | | 1 | | | | | | | |
| **Social Network** | Douban | 2 | 4 | | | | | | 3 | 5 |
| | WeChat | | | | | | | | 1 | 1 |
| | Twitter | | 2 | | | | | | | |
| | Reddit | | | | | | | | | 2 |
| | Pinterest | | 1 | | | | | | 4 | |
| | Flickr | | | | 1 | | | | | |
| | Hike network | | | 1 | | | | | | |
| | BookCrossing | 1 | 1 | 1 | | 1 | | | 2 | 4 |
| **Academic or Book** | DBbook | 1 | 1 | 1 | | | | 1 | | |
| | DBLP | | 1 | 1 | | | | | | 1 |
| | CiteULike | 2 | 1 | 1 | | | | | | |
| | Goodreads | | | | | | | | 1 | 1 |

**Table 7** continued

| Domain | Dataset | Traditional ML | Path-based | Graph embedding | Deep Learning | | | | | |
|---|---|---|---|---|---|---|---|---|---|---|
| | | | | | DNN | AE | AM | DRL | GNN | DHM |
| | User-Tag | 3 | | | | | | | | |
| | Librarything | | | 1 | | | | | 1 | |
| | Youshu | | | | | | | | 1 | |
| | ACL Anthology Network | | | 1 | | | | | | |
| | OHSUMED | | 1 | | | | | | | |
| **POI** | Gowalla | | 1 | 1 | | | 1 | | 4 | |
| | Trip.com+ Facebook+ Twitter | 1 | | | | | | | | |
| | Foursquare | 1 | 1 | 1 | | | 1 | | | |
| | Brightkite | | 1 | | | | | | | |
| | Google Local | | | | | | | | 2 | |
| | Tripadvisor | 1 | | | | | | | | |
| | NUS-MSS | 1 | | | | | | | | |
| **Other** | Educational website and textbooks | | 1 | | | | | | | |
| | MIT AI + CASAS | | | 1 | | | | | | |
| | REDIAL | | | | | | | | | 1 |
| | VizML corpus | | | 1 | | | | | | |
| | Legal recommendation dataset | | | | | | | | | 1 |
| | Yahoo traffic stream | 1 | | | | | | | | |

**Lemei Zhang** is a postdoctoral fellow at the at the Norwegian Research Center for AI Innovation (NorwAI) at NTNU, Norway. Her research topics include recommender system and user modelling. Specifically, she is focusing on applying data mining and machine learning techniques to design effective algorithms to enhance performance of recommender systems in various domains, such as e-commerce, social networks, and so forth.

**Peng Liu** works as a postdoctoral fellow at the Norwegian Research Center for AI Innovation (NorwAI) at NTNU. His research focuses on natural language processing and recommender systems. His primary interests lie in the areas of language modelling, sentiment analysis, topic modeling, lexical semantics, and recommendation algorithms based on data streams and multimodal contexts such as text, image, and so forth.

**Jon Atle Gulla** is a professor of information systems at the Norwegian University of Science and Technology (NTNU) since 2002 and the director of the Norwegian Research Center for AI Innovation. He has a Ph.D. in computer science from 1993 and holds three M.Sc. degrees in computer science, linguistics, and management. His research is on natural language processing and semantics in the context of recommender systems, search engines, and conversational systems.